\documentclass{article}

\usepackage{arxiv}
\usepackage{subcaption}
\usepackage[utf8]{inputenc} 
\usepackage[T1]{fontenc}    
\usepackage{hyperref}       
\usepackage{url}            
\usepackage{booktabs}       
\usepackage{amsfonts}       
\usepackage{nicefrac}       
\usepackage{microtype}      
\usepackage{graphicx}
\usepackage{natbib}
\usepackage{doi}
\usepackage{tcolorbox}
\tcbuselibrary{skins, breakable}
\usepackage{xcolor}
\usepackage{enumitem}
\usepackage{fontawesome5}
\usepackage{multirow}
\usepackage{amsmath}

\title{%
  {\LARGE Letting the neural code speak:}\\[0.4em]
  {\large Automated characterization of \\monkey visual neurons through human language}%
}

\author{
  Vedang Lad$^{1}$ \And
  Katrin Franke$^{1,2}$ \And
  Tamar Rott Shaham$^{3}$ \And
  Surya Ganguli$^{1}$ \AND
  Andreas S. Tolias$^{1}$\thanks{These authors contributed equally as senior authors. Corresponding author  \texttt{vedanglad@stanford.edu}} \And
  Sophia Sanborn$^{1}$\footnotemark[1] \And
  Nikos Karantzas$^{1}$\footnotemark[1] \AND
  \hfill
  \begin{tabular}[t]{c}
    $^{1}$Stanford University \\
    $^{2}$University of Tübingen \\
    $^{3}$Massachusetts Institute of Technology \\
  \end{tabular}
  \hfill
}

\date{}

\renewcommand{\undertitle}{}
\renewcommand{\shorttitle}{\textit{arXiv} Letting the neural code speak}

\hypersetup{
pdftitle={Letting the neural code speak},
pdfsubject={q-bio.NC, q-bio.QM},
pdfauthor={Vedang Lad, Katrin Franke, Tamar Rott Shaham, Surya Ganguli, Andreas Tolias, Sophia Sanborn, Nikos Karantzas},
pdfkeywords={neuroscience, machine learning, interpretability},
}

\begin{document}

\maketitle
\vspace{-8mm}
\begin{center}
\colorbox{purple!85}{\textcolor{white}{\href{https://github.com/enigma-brain/letting-the-neural-code-speak}{\faGithub~\textsf{Code}}}}\hspace{1em}%
\colorbox{purple!85}{\textcolor{white}{\href{https://enigma-brain.github.io/letting-the-neural-code-speak/}{\faGlobe~\textsf{Website}}}}
\end{center}
\vspace{2mm}

\begin{abstract}
Understanding what individual neurons encode is a core question in neuroscience. In primary visual cortex (V1), mathematical models (e.g., Gabor functions) capture neural selectivity, but no comparable framework exists for higher areas. We show that natural language can fill this role: across macaque V1 and V4, the selectivity of most neurons is captured by concise, verifiable semantic descriptions. Using digital twins of V1 and V4, we develop a closed-loop framework that \textit{translates} each neuron's high- and low-activating images into dense captions, generates a \textit{semantic hypothesis} and synthesized images, and \textit{verifies} the hypothesis in silico. Descriptions range from oriented edges and spatial frequency in V1 to conjunctions of form, color, and texture in V4. In V4, images generated from activating and suppressing hypotheses drove 96.1\% of neurons above the 95th and 97.6\% below the 5th percentile of natural-image responses, respectively (vs.\ $\sim$10\% for random images); V1 activation results matched V4, while V1 suppression was less describable in language. Representational similarity analysis reveals partial alignment between neural activity, vision embeddings, and language embeddings, with vision most aligned to neural activity; alignment lost in the text bottleneck is recovered when hypotheses are rendered back into images, showing that linguistic compression is lossy yet semantically faithful. Together, these results show that combining generative models with neural digital twins enables interpretable, testable descriptions of neural function at scale, toward agentic scientific discovery.
\end{abstract}


\section{Introduction}

A central goal of neuroscience is to understand what information individual neurons represent and how their response properties give rise to perception. In early visual cortex, this question has tractable answers: neurons in primary visual cortex (V1) are well described by Gabor functions \citep{jones1987evaluation, daugman1985uncertainty}, capturing orientation and spatial frequency tuning with clear connections to efficient coding principles \citep{olshausen1996emergence, simoncelli2001natural}. Beyond V1, no comparable general framework exists. Neurons in higher visual areas like V4 and inferotemporal cortex exhibit selectivity for curvature \citep[e.g.][]{pasupathy2001shape}, texture \citep[e.g.][]{freeman2013metamers}, faces \citep[e.g.][]{tsao2006stereotyped}, and feature conjunctions \citep[e.g.][]{brincat2004underlying}, but these characterizations have emerged individually from curated stimulus sets designed to test specific hypotheses.

\newpage
Functional ``digital twin'' models have transformed how we probe neural selectivity. Deep networks trained on single-neuron responses to natural stimuli predict activity with high accuracy \citep{walker2019inception, yamins2014performance, cadieu2014deep, Maheswaranathan2023-cp}, and when combined with modern recording technologies \citep{Steinmetz2021-lh, Demas2021-kk}, they enable in-silico screening of millions of stimuli across hundreds or thousands of neurons \citep{Franke2026-qb}. However, prediction does not yield interpretation: identifying images that drive a neuron does not reveal what visual features are encoded by the neuron. Therefore, despite being the most accurate mathematical models of neuronal tuning now available, these deep networks remain opaque. This raises a central challenge: whether neural selectivity, as captured by such models, can be distilled into a compact human-interpretable description.

This gap between predictive accuracy and mechanistic understanding is not unique to neuroscience. Across scientific domains where large-scale measurement has outpaced mechanistic interpretation, such as protein structure prediction \citep{Jumper2021-yv} or weather forecasting \citep{Bi2023-ni}, models achieve remarkable predictive performance while leaving the underlying principles opaque. This reflects broader tension between forms of description: simple mathematical models are interpretable but limited in scope, whereas highly expressive models capture complex phenomena but resist human understanding.

\begin{figure}[t!]
\centering
\includegraphics[width=0.95\linewidth]{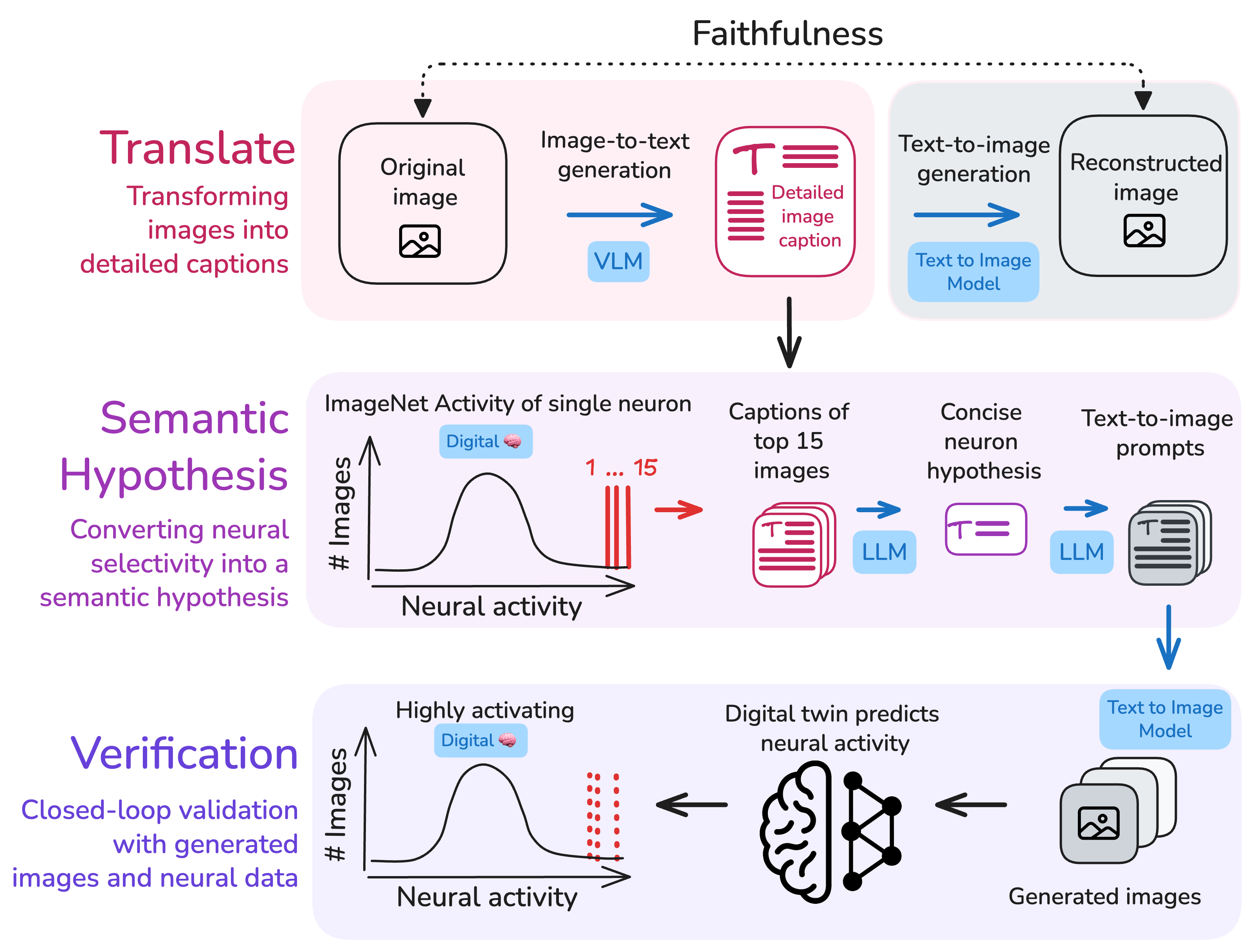}
\caption{\textbf{Framework for translating neural selectivity into interpretable semantic hypotheses.}
    The pipeline consists of three stages: \textbf{Translate:} Each image is converted into a detailed textual description using Gemini 3.0 Pro. To evaluate the fidelity of this image-to-text translation, we regenerate images from the captions using a text-to-image model and quantify correspondence to the original image in an image-similarity space.
    \textbf{Semantic Hypothesis:} For each neuron in our macaque V1 and V4 datasets \citep{Franke2026-qb}, we use a functional digital twin to screen a large-scale naturalistic image database (1.2 million ImageNet images) and identify the stimuli predicted to most strongly drive or suppress it. Gemini 3.0 Pro then analyzes the captions of these extreme response sets to derive a concise, human-interpretable semantic hypothesis describing the neuron's selectivity.
    \textbf{Verification:} We convert the derived hypothesis into prompts for a text-to-image model, generate new images, and use the digital twin to predict neural responses. Agreement between predicted and observed activity provides a closed-loop validation of the neuron’s hypothesized semantic tuning.}
\label{fig:overview}
\end{figure}


Here, we ask whether large language models can bridge this gap, by translating the behavior of complex predictive models into human-interpretable descriptions. In visual cortex, especially at mid to high levels, the features that define neural selectivity (shapes, textures, object parts, and their conjunctions) are features that humans routinely describe in words. We test this idea at two levels. First, at the level of individual neurons, we examine whether language can capture the features that drive and suppress neural activity strongly enough that images regenerated from the description elicit comparable responses to each neuron's preferred and non-preferred natural images. Second, at the population level, we test whether language preserves the relational structure of the neural population code, such that similarity in neural response space is reflected in language embedding space. If so, language would constitute a principled coordinate system for neural selectivity rather than a mere labeling scheme.

We test this directly in macaque V1 (n = 438 neurons) and V4 (n = 205 neurons). For each neuron, we screen 1.2 million ImageNet images through a digital twin to isolate the 15 most- and 15 least-activating stimuli, and use Gemini 3.0 Pro to translate each set into a concise natural-language hypothesis. We then verify each hypothesis in a closed loop against the digital twin: Gemini 3.0 Pro generates prompts describing the visual features that drive or suppress each neuron, Imagen 4.0 synthesizes novel images based on these prompts, and the digital twin measures their predicted responses, with affine transformations to align each stimulus to the preferred position, scale, and orientation of each neuron’s receptive field. The recovered hypotheses span the visual hierarchy, from oriented edges and spatial frequency tuning in V1 to conjunctions of form, color, and texture in V4, and hypothesis-generated images drove $>$96\% of neurons in both V1 and V4 above the 95th percentile of natural image responses; by contrast, randomly drawn ImageNet images reached this threshold for only 10\% of neurons. Suppressive hypotheses drove 97.6\% of V4 neurons below the 5th percentile, but were substantially weaker in V1 (56\%), maybe reflecting the limits of linguistic expressibility for sub-lexical features. Using representational similarity analysis over neural activity, DINOv3 visual embeddings \citep{simeoni2025dinov3}, and Qwen 0.6B language embeddings \citep{zhang2025qwen3}, we further find a partially shared geometric organization across modalities: language not only labels individual neurons but reflects the structure of the neural code.

\section{Related Work}

\paragraph{Automated interpretability in machine learning}

The field of vision model interpretability has evolved from manual visualization-based methods toward increasingly automated approaches. Early work relied on visualizing learned features through activation maximization \citep{zeiler2014visualizing, olah2017feature} or analyzing model responses to curated image datasets \citep{bau2017network, bau2020understanding}. These "Network Dissection" approaches \citep{bau2017network} enabled systematic characterization of neuron selectivity but required pre-defined concept vocabularies and manual curation.

The integration of language models has enabled more flexible semantic interpretation. Initial methods extracted semantic information from visual inputs but still relied on human annotations to label neuron selectivity \citep{mu2020compositional, hernandez2021natural}. More recent approaches automate this process entirely by leveraging vision-language models such as CLIP \citep{oikarinen2023clipdissectautomaticdescriptionneuron, radford2021learning} to generate natural language descriptions of neuron function \citep{hernandez2021natural, gandelsman2023interpreting, bai2025interpret, pmlr-v202-kalibhat23a}. Parallel efforts have extended language-based neuron labeling to large language models themselves, demonstrating that language models can generate and evaluate semantic descriptions of internal units at scale \citep{bills2023language,paulo2025automaticallyinterpretingmillionsfeatures}. However, questions about the reliability of natural language for explainability \citep{huang2023rigorously} have motivated complementary approaches that express representations as interpretable concepts \citep{kim2018interpretability, andreas2017analogs, laina2020quantifying}.

Applying these methods to biological neurons in a neuroscientific setting presents distinct challenges. Unlike artificial networks with well-defined architectures, biological neurons require characterization across diverse response properties, ranging from low-level features in V1 to complex selectivity in V4, and validation through closed-loop experimentation rather than ground-truth labels. Our framework extends recent work exploring agentic hypothesis generation~\citep{shaham2024multimodal, schwettmann2023find} to biological neural data, combining vision-language models with digital twins to enable automated hypothesis generation and verification.

\paragraph{Neural selectivity and single-neuron interpretability}

Early visual areas have long been amenable to compact mathematical description. Neurons in primary visual cortex (V1) are well characterized as oriented, bandpass filters, often modeled with Gabor-like (wavelet-like) functions \citep{jones1987evaluation, daugman1985uncertainty}, connecting naturally to efficient coding principles \citep{olshausen1996emergence, simoncelli2001natural}. Similarly, retinal ganglion cells are modeled as difference-of-Gaussians filters reflecting center-surround antagonism \citep{rodieck1965quantitative}. In higher visual areas, such compact characterizations have proven elusive. Neurons in inferotemporal cortex and V4 exhibit clear selectivity for complex features, including object parts \citep{tanaka1991coding}, curvature \citep{pasupathy2001shape, pasupathy2002population}, faces \citep{tsao2006stereotyped, freiwald2010facepatches}, texture \citep{freeman2013metamers}, and shape conjunctions \citep{brincat2004underlying}, yet no simple mathematical framework has emerged. Cascaded filter models \citep{Movshon2014-vb} successfully capture texture perception \citep{freeman2013metamers} and V4 responses to natural videos \citep{Oliver2024-ec}, but require hand-crafted architectures that do not generalize across diverse feature preferences.

Beyond the question of how to describe selectivity lies the question of whether individual neurons are interpretable at all. Some neurons exhibit remarkably specific tuning, as exemplified by the ``Jennifer Aniston neuron'' that responds selectively across viewpoints and contexts \citep{quiroga2005invariant, quiroga2008human}, yet work in both artificial and biological networks suggests information is predominantly encoded through distributed population codes \citep{kriegeskorte2013representational, zhang2018interpretable}. In deep networks, interpretable neurons represent only a fraction of features \citep{bau2020understanding}, while population-level analyses reveal more robust representations \citep{zhou2018interpretable}, leading to discussions of ``superposition" \citep{elhage2022toymodelssuperposition} and the linear representation hypothesis \citep{park2024linearrepresentationhypothesisgeometry}. However, sampling bias may systematically underestimate single-neuron interpretability: traditional studies test neurons with hundreds to thousands of stimuli \citep{yamins2014performance, cadieu2014deep}, potentially missing optimal configurations. When stimulus diversity increases, neurons that appeared uninterpretable can reveal clear selectivity \citep{bashivan2019neural}. Functional digital twin models prove transformative here, enabling in-silico screening of millions of images \citep{walker2019inception} at scales infeasible in traditional experiments. Our framework leverages digital twins not only to identify optimal stimuli but to translate response patterns into semantic descriptions that can be verified through generative testing.

\paragraph{Language-brain alignment}

A key premise of our approach is that language-derived semantics can meaningfully describe biological neural selectivity. Converging evidence supports alignment between high-level visual cortex and linguistic representations, though most prior work establishes this through predictive modeling rather than testing whether geometric structure is preserved across modalities. \citet{wang2023better} showed that CLIP models predict high-level visual cortex activity more accurately than purely vision-based models, and more broadly, \citet{schrimpf2021neural} demonstrated that transformer language models predict nearly all explainable variance in neural responses to sentences, establishing that next-word prediction architectures capture fundamental aspects of language processing in the brain. \citet{doerig2025highlevel} found that LLM embeddings of scene captions outperform vision models at predicting brain responses, enabling caption reconstruction from activity alone. \citet{bosch2025brain} embedded fMRI activity directly into an LLM's latent space, enabling bidirectional translation between cortical representations and language.

Complementarily, \citet{tang2023semantic} showed that continuous natural language can be reconstructed from non-invasive fMRI recordings, recovering the meaning of perceived and imagined speech, demonstrating that the mapping between neural activity and language is sufficiently structured to support decoding in both directions. \citet{pinier2025large} found alignment between LLM intermediate layers and human frontal potentials during abstract reasoning. However, the nature of this alignment remains debated: \citet{conwell2025rethinking} demonstrated that LLM predictions depend almost entirely on noun information, suggesting brain-language correspondence may reflect shared reference to entities rather than deep structural alignment. \citet{kramer2023features} found that semantic features dominate object memorability, with sensitivity increasing along the visual hierarchy consistent with progressive semantic extraction. These findings connect to a broader debate about whether diverse models converge on similar representations \citep{huh2024platonicrepresentationhypothesis}, with recent work suggesting that this convergence is local and neighborhood-based rather than global \citep{brbic2026aristotelian}, raising the question of whether the brain exhibits analogous structure.

\paragraph{Language-based methods for interpreting neural selectivity}

These findings have motivated methods that use language models to interpret neural data directly. The existence of multimodal neurons in CLIP, units that respond to the same concept whether presented as an image, drawing, or text~\citep{goh2021multimodal}, provides a striking artificial analogue suggesting that single units can carry interpretable, cross-modal semantic content. \citet{luo2024brainscuba} generated fine-grained captions describing visual cortex selectivity, while \citet{Wasserman2025-if} developed an automated framework revealing thousands of visual concepts through natural-language descriptions of fMRI activity. \citet{tilbury2025characterizing} used an LLM-based system to discover orientation tuning equations, and \citet{bai2025interpret} introduced training-free methods for describing artificial neuron selectivity, potentially applicable to biological systems. Taking a complementary approach at the population level, \citet{wang2025migvis} used disentangled latent variable models combined with mutual information-guided diffusion to identify neural population subspaces in macaque IT cortex with clear semantic selectivity for features including object pose and inter-category variation. Our work builds on these foundations by combining vision-language models with digital twins to generate and verify semantic hypotheses about single-neuron selectivity. This closed-loop framework moves beyond correlational alignment toward causal validation and directly tests whether the geometric structure of neural selectivity is preserved across visual and linguistic representations.

\section{Results}
\label{sec:results}

Our framework for interpreting neural selectivity proceeds in three stages (Fig. \ref{fig:overview}). \emph{Translate} converts images into detailed textual descriptions. \emph{Semantic Hypothesis} synthesizes these descriptions into testable predictions about what drives or suppresses individual neurons. We use the term ``semantic hypothesis'' to denote a one-shot natural-language description of a neuron's preferred (or suppressive) visual features. The description is not iteratively refined against neural responses; it is a single-pass summary that is then evaluated through closed-loop generative testing. \emph{Verification} then evaluates these hypotheses by generating novel stimuli and measuring whether neural responses match the predictions from the digital twin model. Together, these stages enable automated, scalable interpretation of neural selectivity through the integration of vision-language models, digital twin neural models, and generative models.

\begin{figure}[b!]
\centering
\captionsetup{position=top}
\caption{\textbf{Translation and faithfulness of image-to-text descriptions.}
    The \emph{Translate} stage of our framework converts input images into detailed captions via Gemini~3.0~Pro and assesses faithfulness by comparing caption-conditioned reconstructions to the originals in DINOv3 embedding space.}
\label{fig:translate}
\begin{subfigure}{\linewidth}
    \centering
    \includegraphics[width=\linewidth]{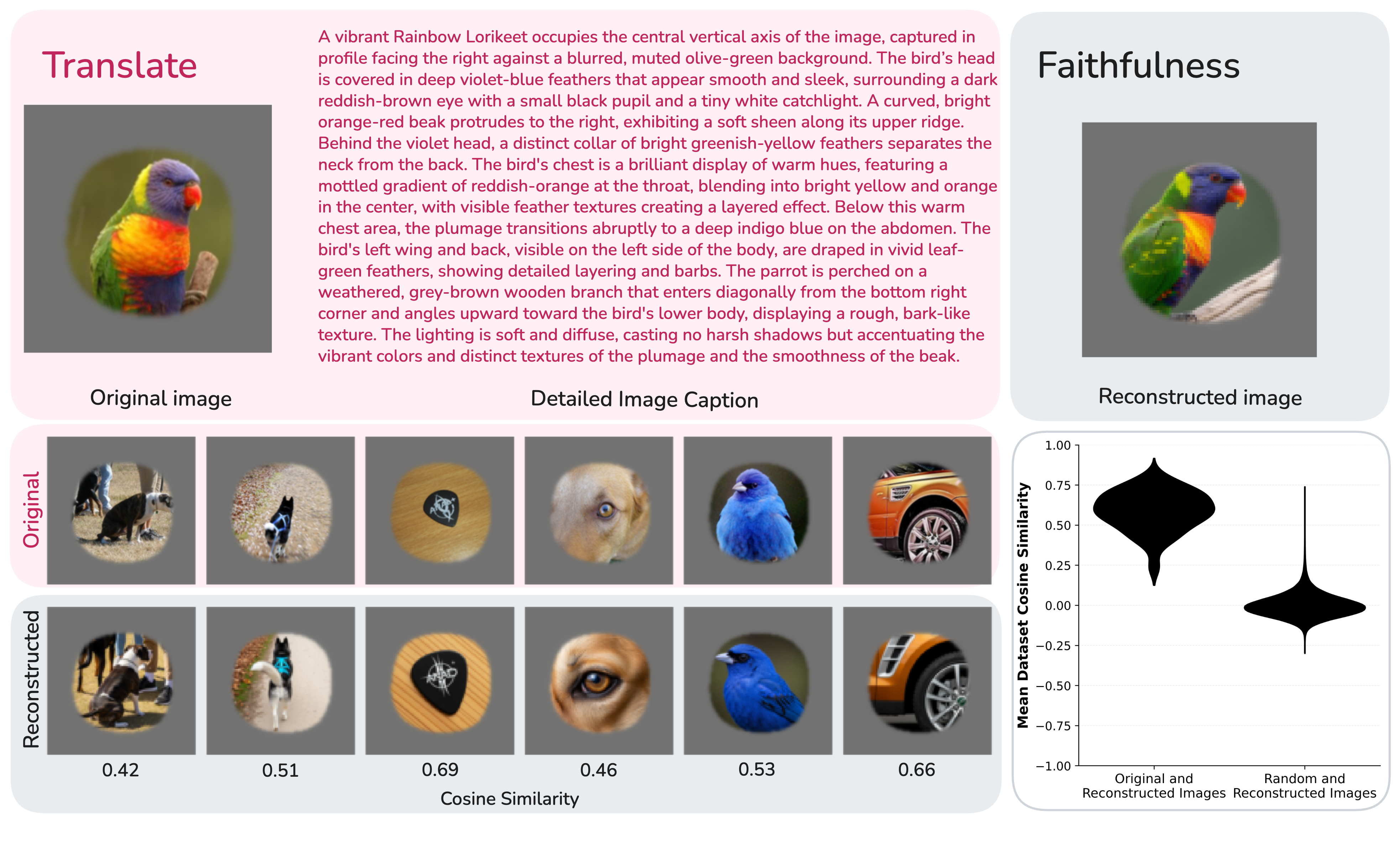}
    \caption{\textbf{Translate: Area V4.} Given an input image (top left), Gemini~3.0~Pro generates a detailed, multi-sentence caption describing the image's visual content (top middle). To assess caption faithfulness, we generate a caption-conditioned reconstruction with a text-to-image model (top right) and quantify similarity to the original in a learned embedding space. Bottom left shows example original and reconstructed images; cosine similarity between DINOv3 embeddings of the original and reconstructed images is indicated below each image pair. Distribution of cosine similarities (violin plots) between each reconstruction and its original image, compared to similarities between the reconstruction and 100 randomly sampled unrelated images.}
    \label{fig:translate_v4}
\end{subfigure}
\end{figure}
\begin{figure}\ContinuedFloat
\begin{subfigure}{\linewidth}
    \centering
    \includegraphics[width=\linewidth]{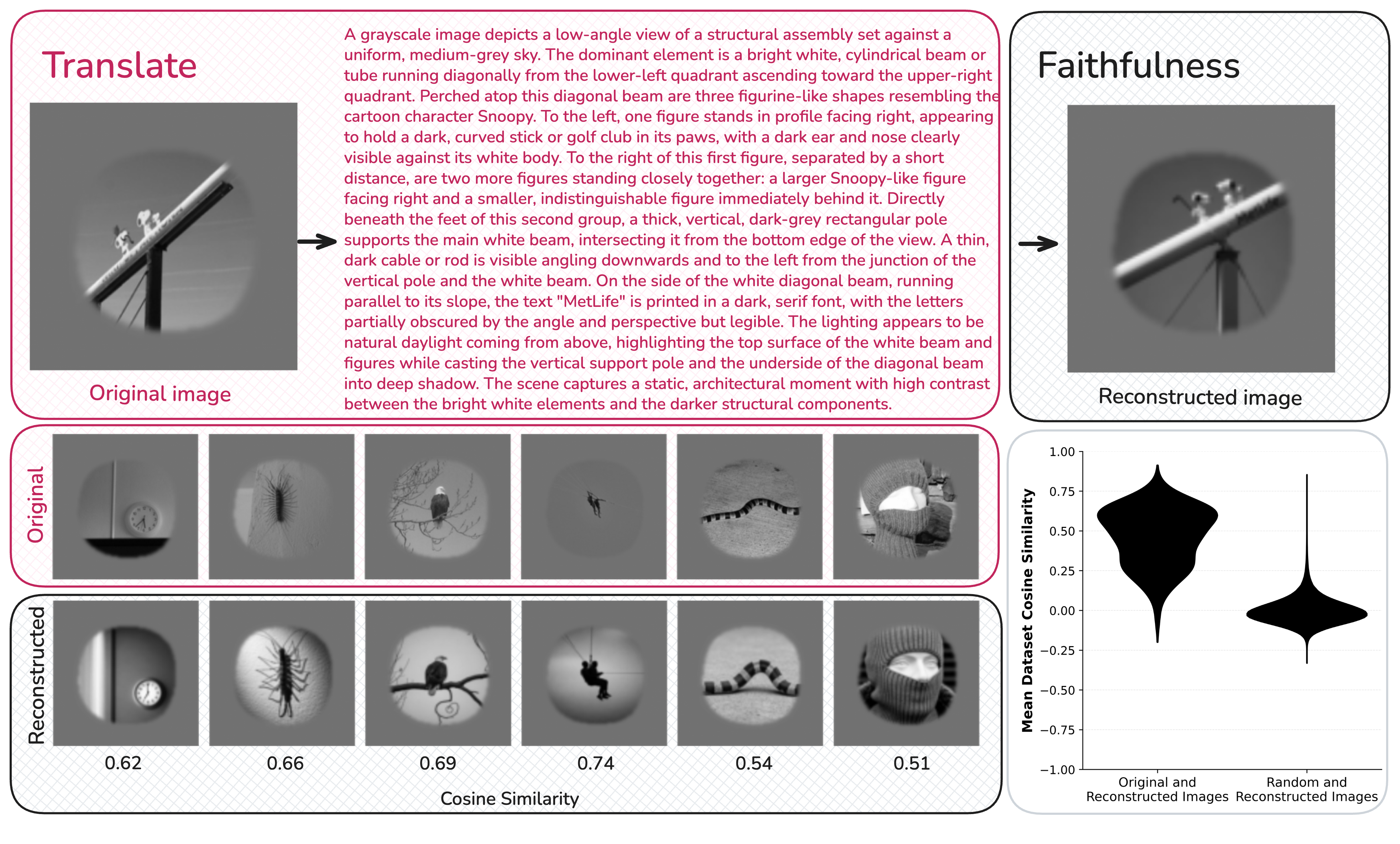}
    \caption{\textbf{Translate: Area V1.} Same analysis as in \textbf{(a)}, applied to the V1 digital twin. V1 stimuli are grayscale rather than RGB due to experimental differences. Reconstructions remain more similar to their source images than to unrelated images (violin plot), confirming that the dense captioning procedure preserves visually relevant information across both grayscale and color image sets.}
    \label{fig:translate_v1}
\end{subfigure}
\end{figure}

\subsection{Translating images into detailed captions validated by reconstruction fidelity}

Here, our goal is to capture single-neuron selectivity in primate visual cortex in the form of human-interpretable semantic descriptions. To this end, we analyze spiking responses of single neurons in macaque V1 and V4, spanning an early and a mid-level stage of the visual stream. We leverage neural predictive models trained on single-neuron spiking data recorded while monkeys viewed natural images during fixation \citep{Franke2026-qb}, and use these functional ``digital twin'' models to identify, for each neuron, images eliciting extreme responses relative to the broader image distribution as a proxy for neural selectivity. The V1 dataset consisted of grayscale images, whereas the V4 dataset used full-color images, reflecting differences in the original experimental protocols.

To translate this selectivity into semantic descriptions, we convert each image into a detailed caption. Routing through text lets us leverage the reasoning capabilities of large language models, which can compare, abstract, and generalize across descriptions in ways not readily available in pixel space (see Section~\ref{sec:discussion}). Specifically, we used Gemini 3.0 Pro to produce dense, multi-sentence descriptions, instructing the model to prioritize visually grounded detail sufficient for reconstruction from text alone (see Methods \ref{sec:methods}). This emphasis on visual fidelity over semantic summarization yielded captions that specify colors, textures, spatial relations, and lighting conditions rather than relying on categorical labels (Figs. \ref{fig:translate_v4} and \ref{fig:translate_v1} top left).

Because the quality of downstream semantic hypotheses depends on the fidelity of this image-to-text mapping, we evaluated whether captions retain visually relevant information using a reconstruction-based check. For each original image, we synthesized a new image from its caption using Imagen 4.0, and quantified similarity in DINOv3 feature space by computing cosine similarity between the original and reconstructed images. We then compared these matched similarities to a null distribution obtained by pairing each reconstruction with randomly sampled images.

Reconstructions showed high similarity to their corresponding source images (Figs. \ref{fig:translate_v4} and \ref{fig:translate_v1} top right and bottom left). Qualitatively, they preserved to varying extents the dominant visual content, including object categorization, pose estimation, color palette, and coarse spatial configuration, despite variability in fine details. For example, a rainbow lorikeet was reconstructed with appropriate plumage coloration and posture, a German shepherd retained breed-typical appearance and positioning, and a car wheel preserved its distinctive orange coloration and overall geometry. Across 100 randomly selected control images, similarity between matched original-reconstruction was substantially higher than similarity between reconstructions and random images (Figs. \ref{fig:translate_v4} and \ref{fig:translate_v1} bottom right). We performed this analysis separately for the V1 and V4 image sets, which differed in whether images were grayscale or full-color as well as in image resolution, and found consistent results in both cases.

These results indicate that our dense captioning procedure produced text descriptions that faithfully preserve visually relevant information needed to characterize neural selectivity. The captions, therefore,  serve as an intermediate, semantically accessible representation for downstream analyses while remaining tightly anchored to the images’ underlying visual structure.

\begin{figure}[b!]
\centering
\captionsetup{position=top}
\caption{\textbf{Deriving semantic hypotheses from neurons in macaque visual cortex.}
    For each V1 and V4 neuron, extreme-response images are identified from a large naturalistic image dataset via a functional digital twin. For neurons with baseline activity, we extract both top- and bottom-activating images and distill each set separately into an excitatory and a suppressive semantic hypothesis; for sparse neurons, we extract only top-activating images and derive a single excitatory hypothesis.}
\label{fig:hypothesis}
\begin{subfigure}{\linewidth}
    \centering
    \includegraphics[width=0.99\linewidth]{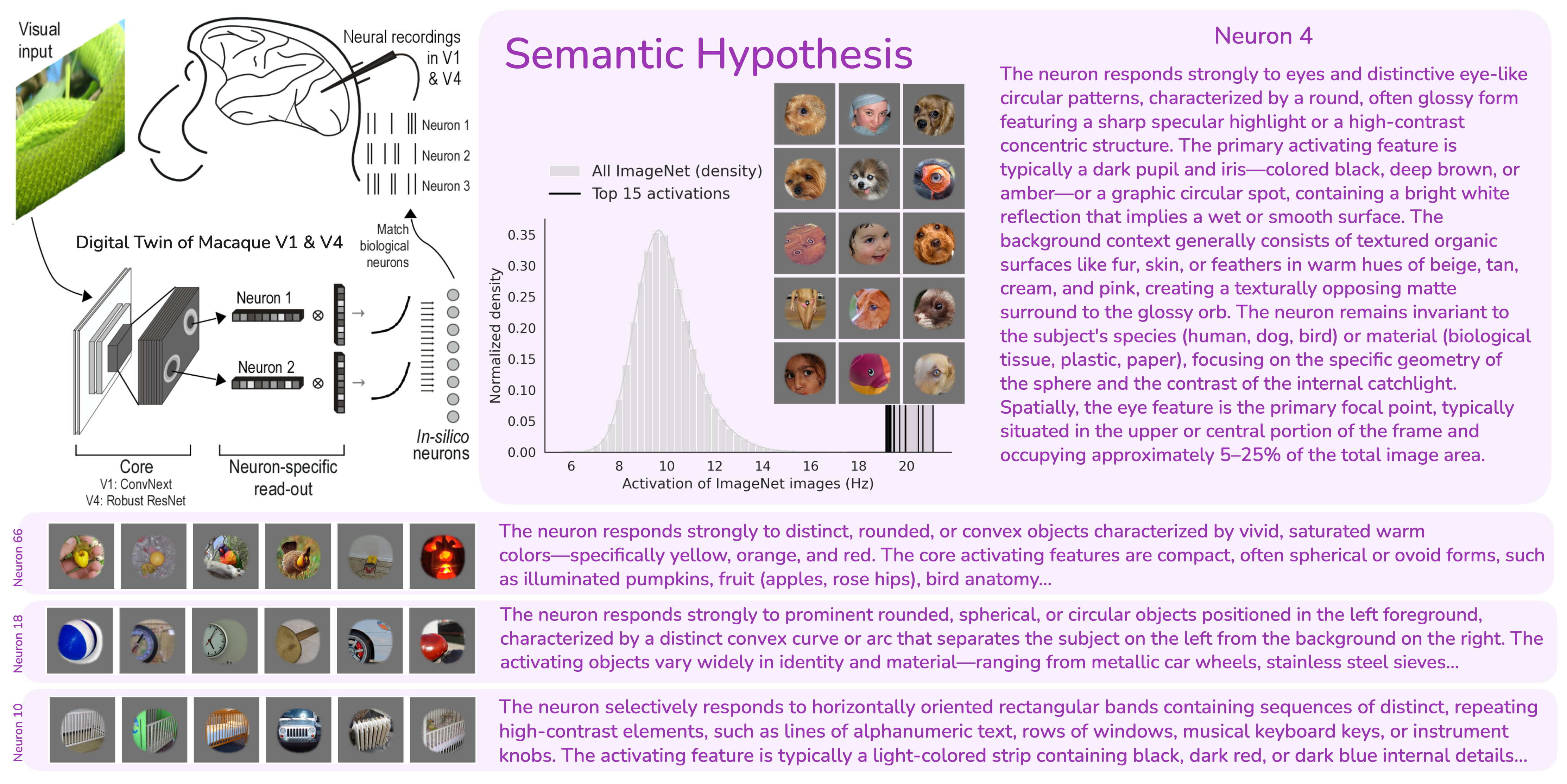}
    \caption{\textbf{Hypothesis: Area V4.} \textbf{Top left:} Functional digital twins of macaque V4 are constructed by training neuron-specific readouts on top of a pretrained CNN backbone, producing in-silico neurons matched to biological neurons. \textbf{Top right:} For a given in-silico neuron, we screen a large naturalistic image dataset and identify the most and least activating images. The distribution of predicted responses across all images (gray) and the top-activating images (black) illustrates the neuron's selectivity. Captions of the top and/or bottom 15 images are condensed into a concise semantic hypothesis; in this example, the neuron is selective for a prominent, high-contrast eye-like feature. \textbf{Bottom:} Additional examples of top-activating images for other neurons, each accompanied by the first two sentences of its derived semantic hypothesis, demonstrating the diversity of interpretable selectivity profiles uncovered by the method.}
    \label{fig:hypothesis_v4}
\end{subfigure}
\end{figure}

\begin{figure}[t!]\ContinuedFloat
\begin{subfigure}{\linewidth}
    \centering
    \includegraphics[width=0.99\linewidth]{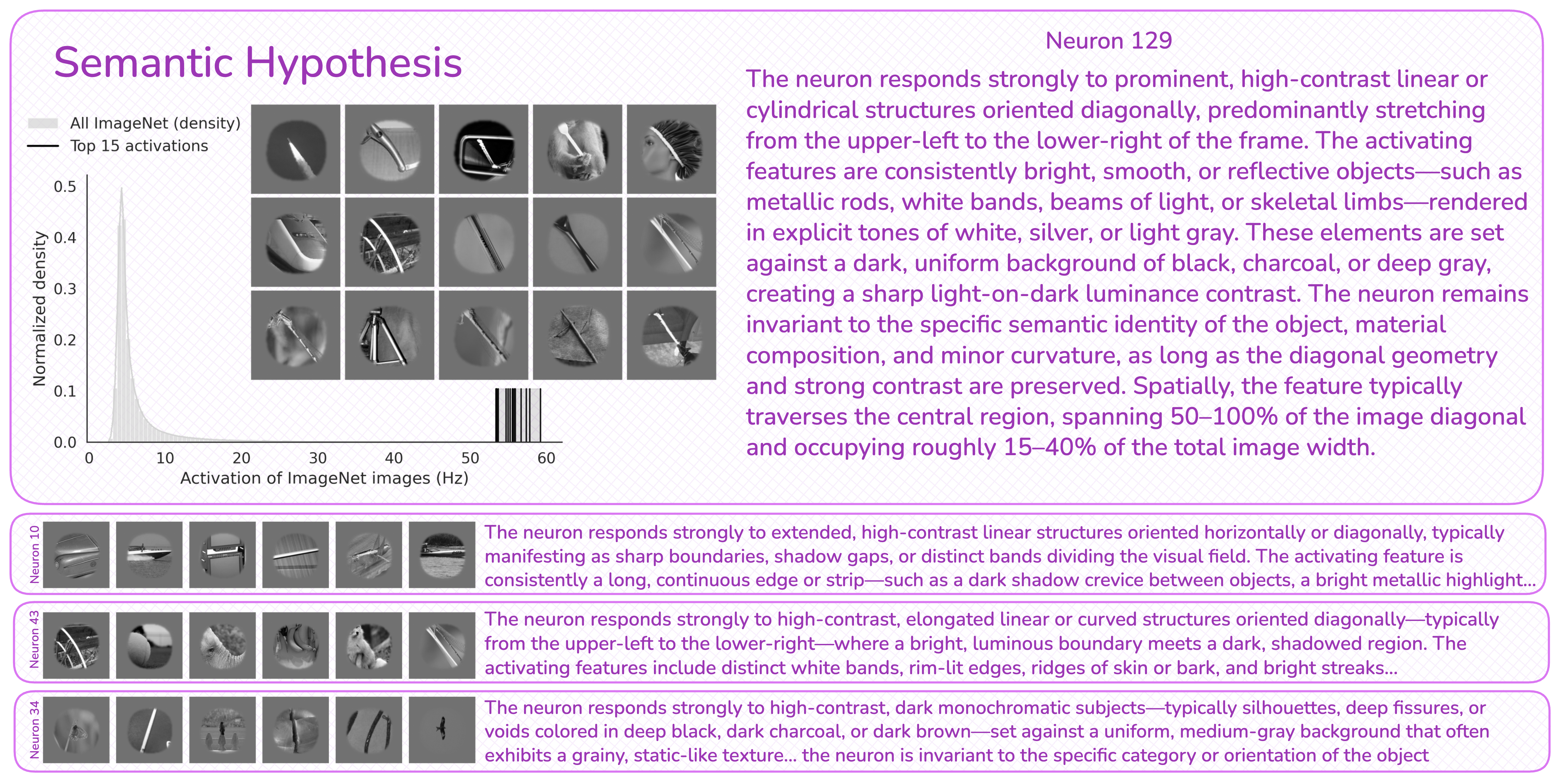}
    \caption{\textbf{Hypothesis: Area V1.} Semantic hypotheses for V1 neurons recover the canonical vocabulary of early visual cortex, namely oriented edges, spatial frequency bands, and contrast polarity, without any domain-specific prior knowledge, providing a ground-truth benchmark for the pipeline. That language recovers these known tuning properties confirms the method's validity before it is applied to less well-characterized areas.}
    \label{fig:hypothesis_v1}
\end{subfigure}
\end{figure}

\subsection{Generating semantic hypotheses of neural selectivity for macaque V1 and V4 neurons}

To map the extreme images of each neuron, we used the digital twins to predict V1 and V4 activity for over one million ImageNet images. For every neuron, this produced a response distribution over images, from which we isolated both the most activating images (MAIs) and the least activating images (LAIs)(Fig. \ref{fig:hypothesis} top middle). We analyzed LAIs only for neurons with high baseline activity, since LAIs carry interpretable information about suppressive selectivity only when the neuron has a non-zero baseline from which it can be suppressed \citep[discussed in][]{Franke2026-qb}. Concretely, we classify a neuron as non-sparse if the skewness of its ordered response distribution across the 1.2 million ImageNet images is below 2; sparsity is a continuum, and this threshold is a pragmatic cutoff rather than a principled boundary.

In line with recent work \citep{Franke2026-qb, Gondur2025-xa}, in both V1 and V4, extreme-response image sets were highly structured: within individual neurons, MAIs were perceptually coherent, and crucially, LAIs of non-sparse neurons were comparably coherent rather than arbitrary weak inputs.

To convert these response patterns into interpretable descriptions, we synthesized semantic hypotheses from the visual content of each neuron’s extreme-response stimuli. For each neuron, we selected the top 15 MAIs and bottom 15 LAIs identified by the digital twin and retrieved their dense captions generated during the \textit{Translate} stage. We then prompted a language model to analyze each caption set and produce a concise hypothesis describing the neuron's selectivity at each end of its activation range, identifying features associated with strong activation or suppression, specifying dimensions of invariance, and summarizing spatial properties such as orientation and scale. This procedure compressed high-dimensional response statistics into a compact, human-interpretable description of the visual structure that reliably resulted in activation or suppression.

The hypothesis-generation prompt was developed iteratively on a small set of held-out V4 neurons. We did not perform a systematic prompt-sensitivity analysis; the reported results are conditional on the final prompt template (Appendix A.1).

Across the population, this procedure revealed a clear progression in semantic complexity. Hypotheses for V1 neurons primarily emphasized low-level features such as orientation, spatial frequency, and contrast polarity. In contrast, hypotheses for V4 neurons captured more complex conjunctions of form, color, and texture. Together, these results show that semantic hypothesis generation yields an interpretable and scalable summary of neural selectivity across multiple stages of the visual hierarchy, setting the stage for in-silico closed-loop verification through generative testing.

\begin{figure}[t!]
\centering
\includegraphics[width=1\linewidth]{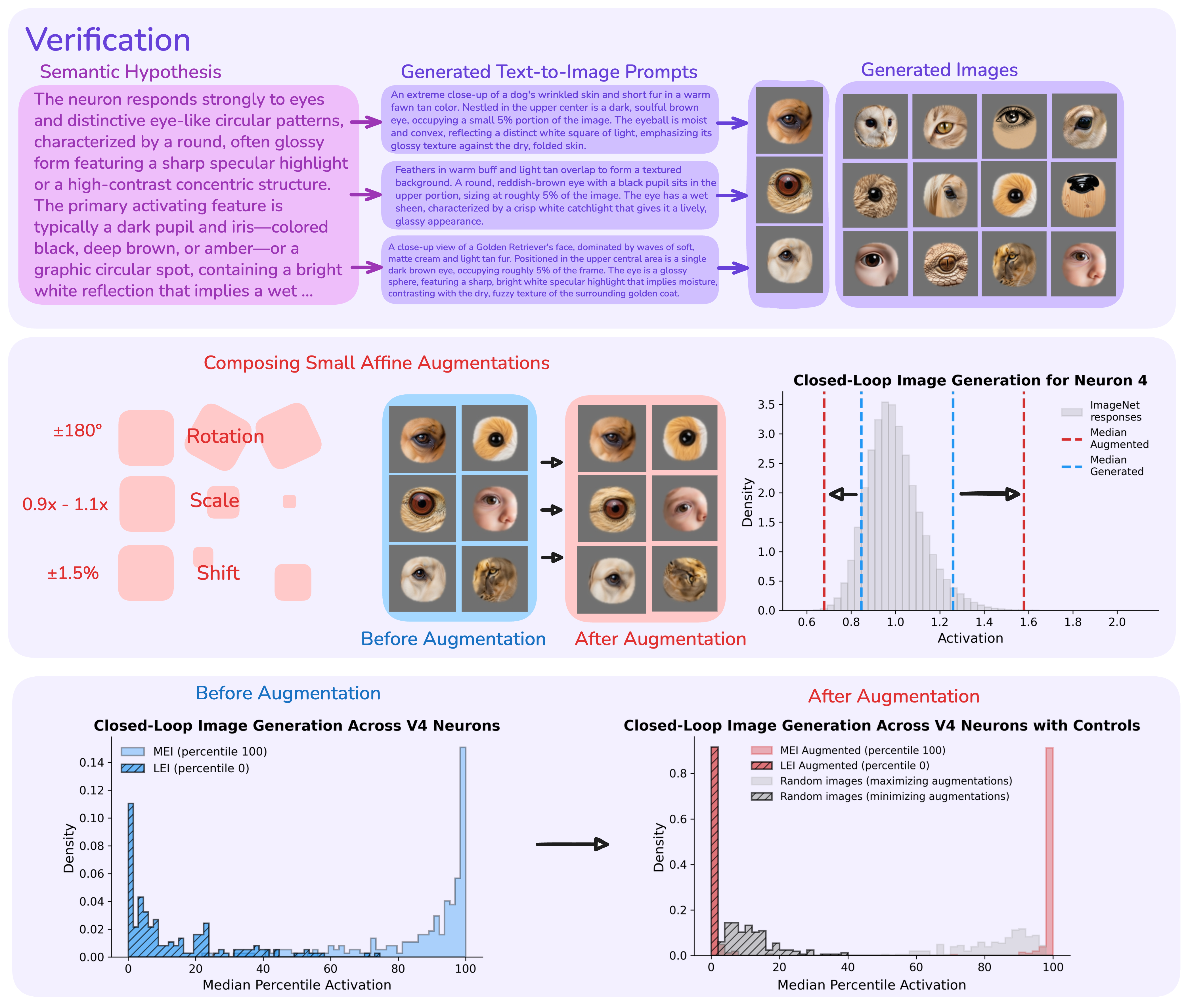}
    \caption{\textbf{Area V4: Closed-loop verification of semantic hypotheses using generative stimuli and spatial optimization.}
    \textbf{Top:} A generated semantic hypothesis for an example V4 neuron is expanded into multiple diverse text prompts, which are then rendered into novel images using a text-to-image model. These generated images resemble the neuron’s most-activating natural images, capturing core feature conjunctions such as round, glossy, eye-like structures with high-contrast specular highlights embedded in organic textures.
    \textbf{Middle:} Response distribution for an example neuron, showing that both generated and affine-optimized images shift activation to the extremes of the ImageNet response distribution.
    \textbf{Bottom} Distribution of neural responses to hypothesis-generated images across the V4 population, expressed as median percentile activity relative to natural images. Generated images already elicit extreme responses, and systematic spatial optimization via small affine transformations (rotation, scale, and translation) further increases activation.
    \textbf{Bottom right (gray):} Control analysis applying identical affine transformations to random images. Unlike hypothesis-generated stimuli, random images show smaller improvement under either maximizing or minimizing augmentations, demonstrating that semantic content, not spatial optimization alone, is necessary for strong activation. }
    \label{fig:verification}
\end{figure}

\subsection{Closing the loop: Verifying semantic hypotheses through generative testing}

Having generated semantic hypotheses for each neuron, we sought to verify whether these descriptions capture sufficient information to predict neural selectivity. A valid hypothesis should enable the generation of novel stimuli that drive the neuron into the extreme tails of its natural-image response distribution. We note that this is a weaker criterion than full equivalence to the original top-15 MAIs or bottom-15 LAIs, which occupy a small region within those tails; we return to this distinction in Section~\ref{sec:limitations}. We implemented this verification through an in-silico closed-loop procedure combining text-to-image generation with subsequent spatial optimization (Fig.~\ref{fig:overview} bottom row).

For each neuron, we converted its semantic hypothesis into multiple image prompts, instructing a language model to generate diverse textual descriptions consistent with the hypothesized selectivity. These prompts were then rendered into images using a text-to-image model, producing novel stimuli (Fig.~\ref{fig:verification} top row). Critically, this procedure tested the semantic content of our hypotheses: if the generated images activated the neuron in extreme response percentiles relative to natural images, the hypothesis captured the relevant visual features; if not, the description was incomplete or inaccurate.

\begin{figure}[t!]
\centering
\includegraphics[width=1\linewidth]{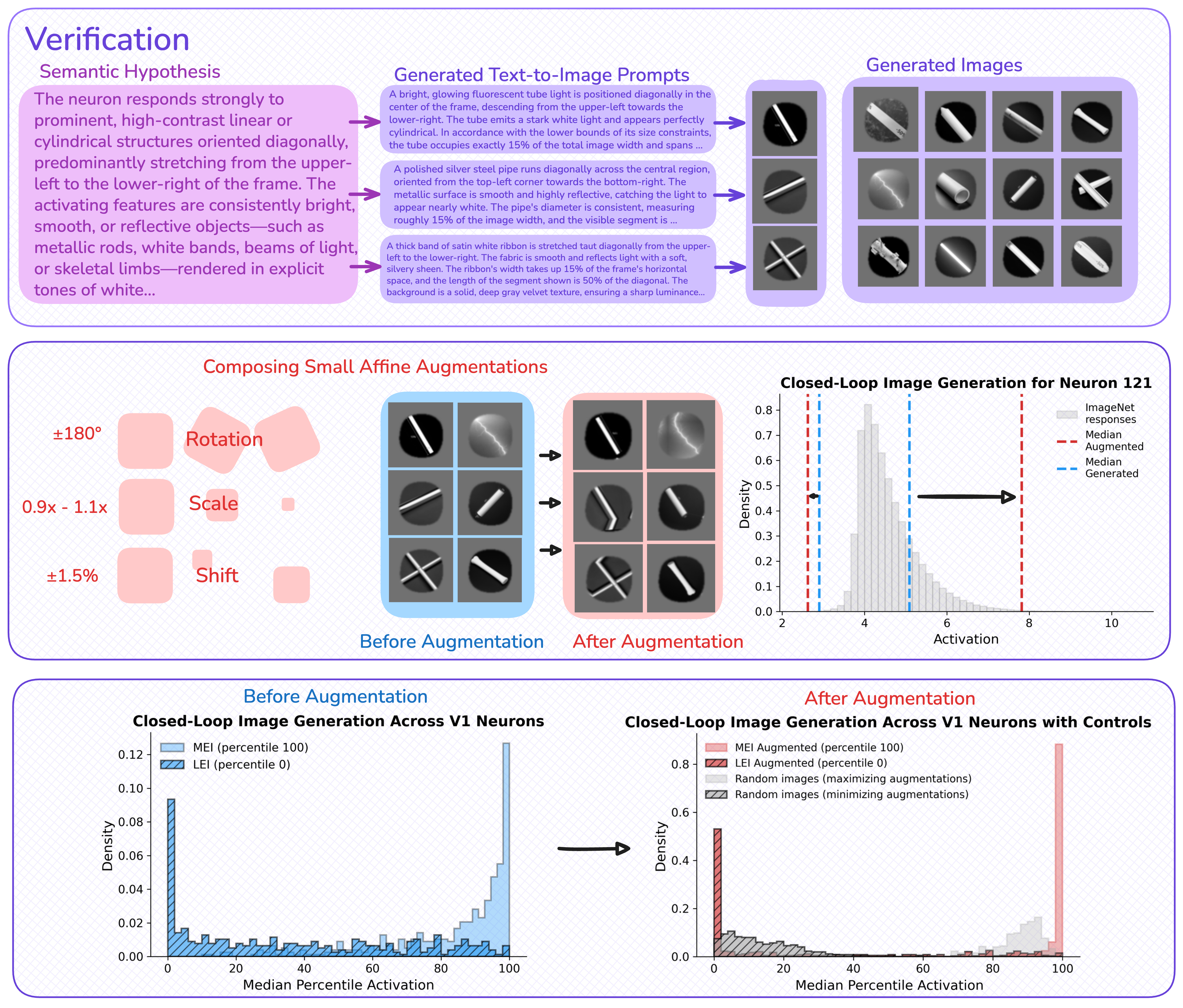}
    \caption{\textbf{Area V1: Closed-loop verification of semantic hypotheses using generative stimuli and spatial optimization.}
    Semantic hypotheses successfully generate stimuli that drive neurons above the random baseline, confirming that the pipeline generalizes across the visual hierarchy. The smaller gain from spatial optimization relative to V4 quantifies the expected gradient: language is a coarser coordinate system for sub-lexical properties such as precise orientation and spatial frequency than for the richer conjunctive features of V4, establishing a principled bound on linguistic expressibility at different cortical stages.}
    \label{fig:verification_1}
\end{figure}

Initial verification revealed that hypothesis-generated images qualitatively resembled the original extreme image sets, capturing core feature conjunctions identified in the semantic hypotheses. For example, a neuron selective for eye-like structures embedded in skin textures produced generated images with similar round, glossy forms featuring specular highlights against organic backgrounds (Fig.~\ref{fig:verification} top right). However, generated images sometimes failed to capture particular aspects of a neuron's tuning; for instance, the hypothesis correctly identified the eye-like feature, but generated images did not fully capture the neuron's invariance to background color. Despite these limitations, hypothesis-generated images resulted in extreme responses relative to the natural image distribution (Fig.~\ref{fig:verification} bottom left).

While semantic hypotheses captured the relevant visual features, vision-language models are less reliable at specifying \textit{where} features should appear within the receptive field, at what scale, or at what orientation (see Section~\ref{sec:discussion}). To address this, we introduced a spatial optimization stage using affine transformations (Fig.~\ref{fig:verification} middle left). For each hypothesis-generated image, we systematically searched over a range of rotations, scales, and translations, to identify the spatial configuration that maximized (or minimized) predicted activation. This two-stage procedure, semantic generation followed by spatial optimization, separated the contributions of feature content from spatial arrangement. To control for the possibility that spatial optimization alone could produce extreme activations regardless of semantic content, we applied identical affine transformations to randomly selected images. Random images showed significantly weaker response shifts (Fig.~\ref{fig:verification} middle-right), confirming that the semantic content of the generated images is important for achieving activation around the extremes of each neuron.

\begin{table}[h!]
\centering
\small
\begin{tabular}{@{}llcccc@{}}
\toprule
\textbf{Area} & \textbf{Condition} & \textbf{Threshold} & \textbf{$n$} & \textbf{Semantic (\%)} & \textbf{Null (\%)} \\
\midrule
\multirow{6}{*}{V4}
  & \multirow{3}{*}{Excitatory (MAI)}
    & $>$90th & 205 & 99.5 & 33.2 \\
  & & $>$95th & 205 & 96.1 & 8.8 \\
  & & $>$99th & 205 & 84.4 & 0.0 \\
\cmidrule{2-6}
  & \multirow{3}{*}{Suppressive (LAI)}
    & $<$10th & 166 & 99.4 & 45.2 \\
  & & $<$5th & 166 & 97.6 & 13.3 \\
  & & $<$1st & 166 & 78.9 & 0.0 \\
\midrule
\multirow{6}{*}{V1}
  & \multirow{3}{*}{Excitatory (MAI)}
    & $>$90th & 438 & 98.4 & 43.1 \\
  & & $>$95th & 438 & 96.1 & 8.1 \\
  & & $>$99th & 438 & 80.6 & 0.7 \\
\cmidrule{2-6}
  & \multirow{3}{*}{Suppressive (LAI)}
    & $<$10th & 343 & 59.5 & 43.8 \\
  & & $<$5th & 343 & 56.6 & 23.5 \\
  & & $<$1st & 343 & 47.8 & 4.0 \\
\bottomrule
\end{tabular}
\caption{\textbf{Verification of semantic hypotheses across V1 and V4.} Percentage of neurons whose best hypothesis-generated image (after affine optimization) exceeded the given response percentile threshold. The ``Null'' column indicates the fraction for which augmentation of random stimuli also reached this threshold, providing a baseline. Non-sparse neurons ($n_{\text{V4}}=166$, $n_{\text{V1}}=343$) are evaluated for both excitatory and suppressive hypotheses; sparse neurons ($n_{\text{V4}}=39$, $n_{\text{V1}}=95$) contribute to excitatory totals only.}
\label{tab:verification}
\end{table}

Table~\ref{tab:verification} quantifies verification performance against a random-augmentation null baseline. Reported percentages are post-augmentation; the contribution of generative content versus affine search alone can be read from the comparison with the random-image null column, which applies the same affine search to images bearing no specific relationship to the neuron. The gap between the two columns is therefore attributable to the semantic content of the hypothesis rather than to the spatial search itself. In V4, excitatory hypotheses drove 96.1\% of neurons above the 95th percentile (null: 8.8\%) and 84.4\% above the 99th (null: 0.0\%). Suppressive hypotheses were equally effective: 97.6\% of non-sparse neurons fell below the 5th percentile (null: 13.3\%) and 78.9\% below the 1st (null: 0.0\%). In V1, excitatory performance matched V4, with 96.1\% above the 95th percentile (null: 8.1\%) and 80.6\% above the 99th (null: 0.7\%). Suppressive hypotheses were less effective in V1, reaching 56.6\% below the 5th percentile versus 97.6\% in V4. This rate is more than double the 23.5\% null baseline, indicating that language captures some suppressive structure in V1, but it leaves 43\% of non-sparse V1 neurons without an effective suppressive description. We discuss possible reasons for this in Section~\ref{sec:limitations}.

\subsection{Cross-modal alignment of neural activity with visual and semantic representations}
\label{sec:rsa}

So far, we have shown that natural language can describe the selectivity of individual neurons in V1 and V4 with high fidelity: concise semantic hypotheses generate novel stimuli that are predicted by the digital twin model to reliably drive neurons to the extremes of their natural response distributions. This establishes a pointwise correspondence: one description per response extreme of a neuron. But neural codes are organized, and neurons with related selectivity occupy related positions in population space. If language truly captures neural selectivity, it should preserve this \emph{relational} structure, not merely label neurons one at a time.

To probe relational structure, we describe each neuron by the images that maximally activate its in-silico counterpart, in line with extensive prior work characterizing neurons through their preferred stimuli. We then ask whether neuron-to-neuron similarity is consistent across representational spaces that describe each neuron's most-activating images at successive levels of abstraction.

We embedded each neuron's most-activating image set into six such spaces: (1)~neural population activity evoked by these images; (2)~visual embeddings from DINOv3 \citep{simeoni2025dinov3}; (3)~Qwen3 0.6B language embeddings \citep{zhang2025qwen3} of dense image captions; (4)~language embeddings of concise semantic hypotheses; (5)~DINOv3 embeddings of images regenerated from those hypotheses; and (6)~neural population responses to the same regenerated images. The first four trace the original stimulus-to-description pathway; the last two close the loop, testing whether images resynthesized from semantic hypotheses recover the original alignment.

\begin{figure}[t!]
\centering
\captionsetup{font=footnotesize}
\captionsetup{position=top}
\caption{\textbf{Cross-modal alignment of neural activity with visual and semantic representations.} Each neuron's activating image sets are embedded in six representational spaces and compared via RSA. Language preserves the geometric structure of neural selectivity at the level of highly activating images. Hypothesis-generated images, derived entirely from semantic hypothesis text, recover and in some cases exceed the cross-space alignment of the hypotheses themselves, demonstrating that text-to-image models re-expand compressed linguistic descriptions back into visually and neurally meaningful representations.}
\label{fig:alignment}
\begin{subfigure}{\linewidth}
    \centering
    \includegraphics[width=0.95\linewidth]{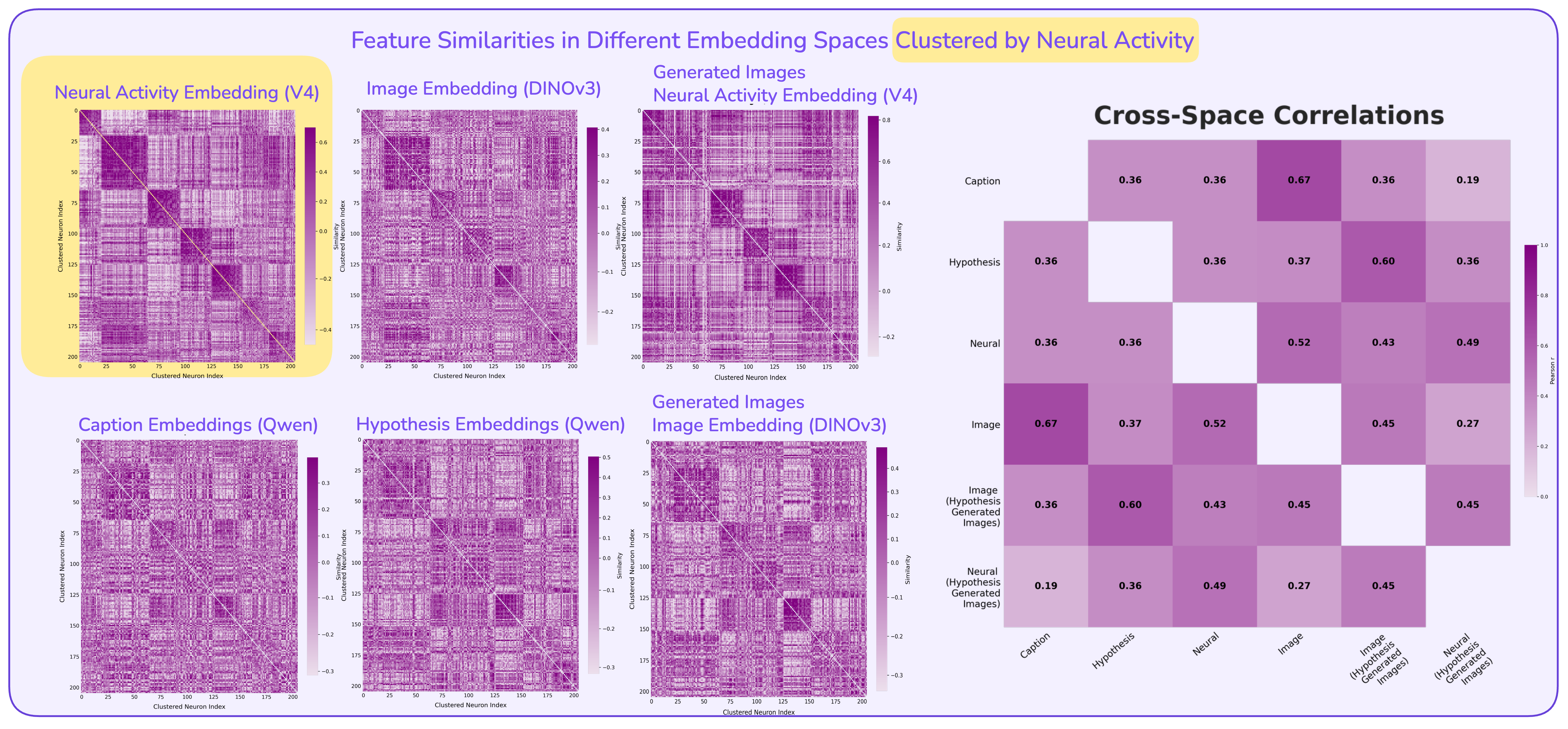}
    \caption{\textbf{Area V4.} \textbf{Left:} Representational similarity matrices (RSMs) across embedding spaces, all ordered by neural similarity computed on a held-out subset of images (see Methods~\ref{sec:rsa_methods}). Shared block structure indicates that clusters defined by population activity are preserved across visual, language, and generative spaces. \textbf{Right:} Cross-space correlations (Pearson $r$, upper triangles); all are far above a held-out label-permutation null ($p < 10^{-4}$; null 95th percentiles $\le 0.020$).  Image and caption embeddings are most aligned ($r = 0.67$). Neural activity aligns strongly with image embeddings ($r = 0.52$) and more modestly with captions ($r = 0.36$). Hypothesis-generated images, synthesized from text descriptions, consistently align with other spaces as well as or better than the text itself (caption: $0.36$ vs.\ $0.36$; neural: $0.43$ vs.\ $0.36$), while remaining most similar to their source hypotheses ($r = 0.60$). Neural responses to generated images preserve the original structure ($r = 0.49$), indicating that the translate-hypothesize-generate loop retains neurally relevant geometry.}
    \label{fig:alignment_v4}
\end{subfigure}

\begin{subfigure}{\linewidth}
    \centering
    \includegraphics[width=0.95\linewidth]{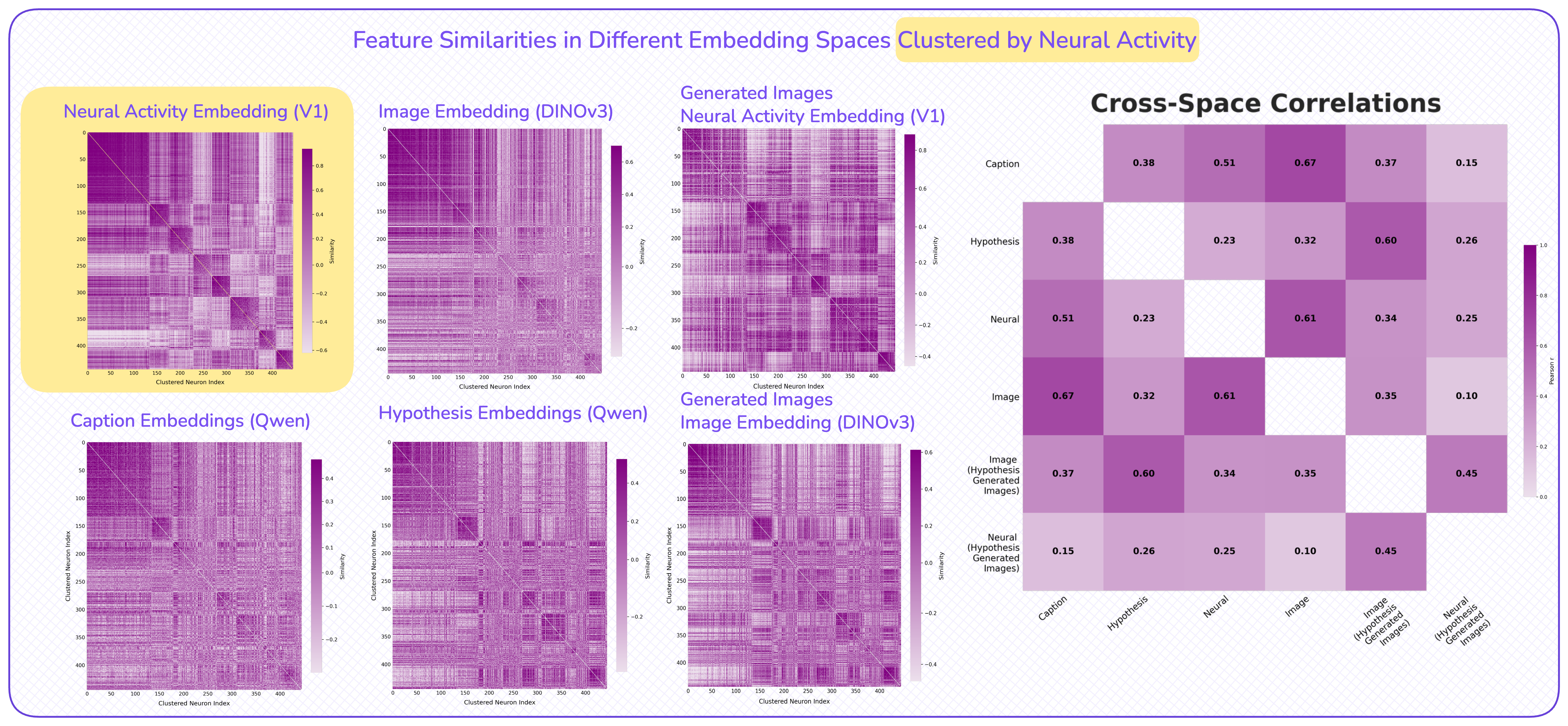}
    \caption{\textbf{Area V1.} The same cross-modal structure is present in V1, with all cross-space correlations above the held-out label-permutation null ($p < 10^{-4}$; null 95th percentiles $\le 0.032$). Image--caption alignment remains high ($r = 0.67$), and neural activity aligns with image ($r = 0.62$) and caption embeddings ($r = 0.51$). Generated images again match or exceed the alignment of hypothesis text (caption: $0.37$ vs.\ $0.38$; neural: $0.34$ vs.\ $0.23$). However, neural responses to generated images align less strongly with the original activity ($r = 0.25$ vs.\ $0.49$ in V4), consistent with the difficulty of capturing low-level features (e.g., orientation, spatial frequency) through text-to-image generation.}
    \label{fig:alignment_v1}
\end{subfigure}
\end{figure}

We quantified alignment by representational similarity analysis (RSA)~\citep{kriegeskorte2008representational}, taking the Pearson correlation between the upper triangles of the neuron-by-neuron similarity matrices for every pair of spaces (Fig.~\ref{fig:alignment}). To guard against alignment that is inherited trivially from the ordering used to build the matrices, we adopt a held-out construction: we cluster neurons in neural-activity space using only seven of the fifteen most-activating images per neuron, and then compute every cross-space correlation on the matrices built from the remaining eight images, ordered by the held-out neural clustering. Significance is assessed against a label-permutation null in which row and column labels of one matrix are jointly permuted 10{,}000 times, preserving its internal geometric structure (Methods~\ref{sec:rsa_methods}). In V4, all six matrices exhibited a consistent block structure, and every cross-space correlation was far above its permutation null (all $p < 10^{-4}$; null 95th percentiles $\le 0.020$). Image and dense-caption embeddings were the most strongly coupled ($r = 0.67$), and neural activity aligned substantially with image embeddings ($r = 0.52$). Neural responses to hypothesis-generated images, i.e. novel stimuli synthesized from semantic text using a text-to-image model, aligned with the original neural activity at $r = 0.49$, consistent with the full translate--hypothesize--generate loop preserving neurally relevant selectivity structure.

V1 exhibited the same six-way alignment pattern (Fig.~\ref{fig:alignment_v1}), with every cross-space correlation again far above its permutation null ($p < 10^{-4}$; null 95th percentiles $\le 0.032$), suggesting that the language--neural geometric correspondence extends across the visual hierarchy. The loop closure was, however, weaker: neural responses to hypothesis-generated images correlated with the original activity at only $r = 0.25$ (vs.\ $r = 0.49$ in V4), which may reflect the difficulty of rendering sub-lexical V1 features through text-to-image generation. This attenuation may trace the boundary of linguistic expressibility from early to mid-level visual cortex.

\begin{figure}[t!]
\centering
\includegraphics[width=0.99\linewidth]{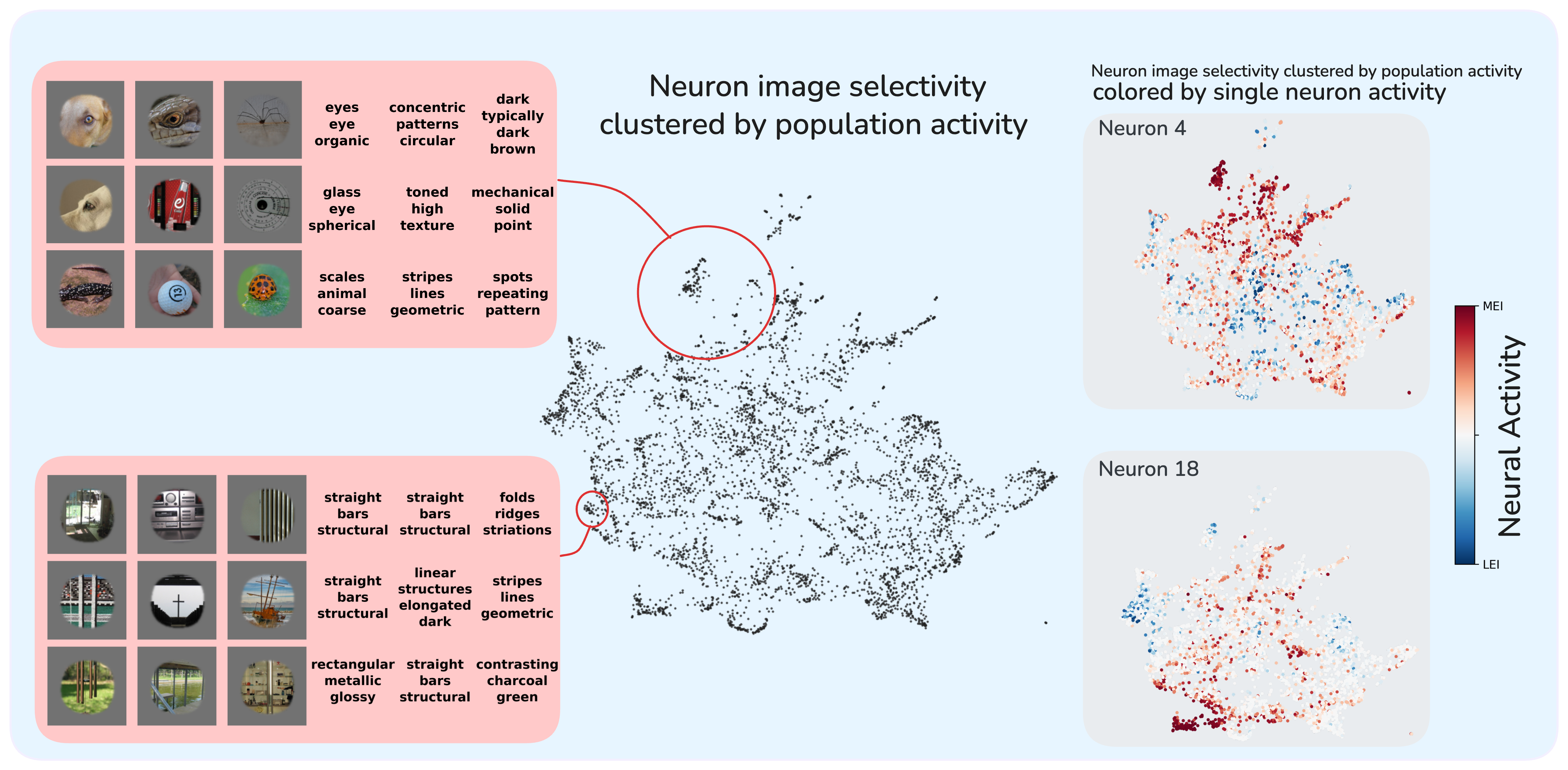}
\caption{\textbf{Semantic structure of neural selectivity revealed through population activity clustering.}
    \textbf{Left:} UMAP embedding of V4 neurons clustered by population activity similarity, annotated with nouns and adjectives extracted from the first sentence of each neuron's semantic hypothesis. Large-scale neighborhoods exhibit smooth transitions in both visual content and descriptive language, from eyes and circular organic patterns to geometric textures and repeating structures, while small neighborhoods contain neurons with nearly identical descriptive features. \textbf{Right:} The same UMAP embedding colored by single-neuron response magnitude for two example neurons, showing that individual neurons tile localized, semantically coherent regions of population activity space with smoothly varying activation.}
\label{fig:umap}
\end{figure}

Together, these results indicate that neural, visual, and linguistic representations share partially overlapping geometric structure at the level of individual neurons, with vision embeddings the most strongly coupled to neural activity and language embeddings consistently weaker. This shared structure survives the cycle of linguistic compression and generative re-expansion in V4, but is attenuated in V1, where sub-lexical features likely tax both the captioning and image-generation steps. We illustrate the shared structure in V4 (Fig.~\ref{fig:umap}) by projecting images into two dimensions via UMAP over V4 population activity, with each point labeled by key nouns and adjectives from its caption. Qualitatively, neighborhoods in this embedding contain images whose captions share descriptive vocabulary, and single-neuron activation varies smoothly across the embedding, consistent with nearby points in population space also being nearby in linguistic space.

\section{Discussion}
\label{sec:discussion}

\subsection{Insights into neural selectivity from semantic descriptions}

For the majority of neurons in macaque V1 and V4, natural language descriptions of response extremes were sufficient to generate novel images whose predicted responses landed in the extreme tails of the natural-image response distribution. This establishes that hypothesis-generated stimuli reach the same tails as the original MAIs and LAIs, though not necessarily the same exact activation levels as the original top-15 or bottom-15 images. This implies that single-neuron selectivity is, in most cases, expressible as semantic description of visual features rather than only as a pattern in a high-dimensional, non-interpretable feature space.

Whether this is surprising depends on one's prior assumptions about the unit of neural coding. A long-standing debate concerns the degree to which individual neurons carry interpretable content versus acting as mixed-selective units whose meaning only emerges at the population level \citep{Rigotti2013-hn, quiroga2005invariant, kriegeskorte2013representational, Thorpe1989-av}. An analogous tension exists in artificial neural networks, between superposition \citep{elhage2022toymodelssuperposition} and single-unit interpretability \citep{Colin2024-nr, bau2020understanding}. Our results do not resolve this debate, but they add quantitative weight to the view that single-neuron interpretability is more prevalent than classically appreciated, particularly when the explored stimulus space is sufficiently large \citep{Franke2026-qb, bashivan2019neural}.

The semantic descriptions we recover for V1 and V4 do not reveal qualitatively new tuning properties. Hypotheses for V1 neurons emphasize oriented edges, spatial frequency, and contrast polarity; those for V4 reflect conjunctions of curvature, texture, and color, precisely the features documented by decades of targeted electrophysiology \citep{jones1987evaluation, daugman1985uncertainty, pasupathy2001shape, freiwald2010facepatches, brincat2004underlying}. Crucially, the pipeline recovers these known V1 properties without any injected prior knowledge about visual neuroscience: the same procedure that discovers eye-like selectivity in V4 \citep{Franke2026-qb, Willeke2026-dq} also rediscovers orientation tuning in V1. This constitutes an internal positive control: the method passes the test where ground truth is known, before being applied where it is not.

The value of the approach lies precisely in this independence from prior knowledge. Advances in large-scale recording technologies are expanding access to cortical areas that lack established stimulus vocabularies \citep{Steinmetz2021-lh, Demas2021-kk}. A framework that operates on arbitrary responses to naturalistic images and returns testable semantic descriptions is well-suited for discovery in such regions. Extending the framework into the spatiotemporal domain, characterizing neurons in terms of dynamic features and motion patterns, would open an interpretability window even in well-studied areas where the temporal structure of selectivity remains difficult to summarize compactly.

\subsection{Language as a geometric coordinate system for neural selectivity}

The RSA results go beyond showing that language is a useful tool for describing individual neurons. Representational similarity analysis is one linear measure of the geometry of neural population responses: it asks whether the pattern of neuron-to-neuron similarities computed from activity is preserved when the same neurons are described in another space. Under this measure, we find that neuron-to-neuron similarities computed from neural activity are preserved, partially, when the same neurons are described in visual feature space and in language embedding space. This partial cross-modal alignment connects to a broader debate about whether diverse models converge on similar representations of the world \citep{huh2024platonicrepresentationhypothesis}, with recent work suggesting that the convergence is local and neighborhood-based rather than global \citep{brbic2026aristotelian}; our results extend this question to biological neural activity, where we similarly find local rather than complete alignment between neural, visual, and linguistic representations. The partial preservation we observe is not guaranteed: a weaker alternative, in which language accurately labels individual neurons but does not preserve neuron-to-neuron similarities, would be fully compatible with our single-neuron verification results and would leave the RSA correlations at chance. The non-trivial magnitude of these correlations, together with their separation from a held-out label-permutation null (all $p < 10^{-4}$; null 95th percentiles $\le 0.032$ across V1 and V4), indicates that this weaker regime does not hold.

The fact that even concise hypothesis embeddings maintain above-chance RSA alignment with neural activity (despite compressing each neuron's selectivity into a short paragraph) suggests that the hypothesis generation procedure extracts features relevant to neural selectivity, rather than surface properties of the captions that are uncorrelated with the neural code, e.g., caption length, syntactic style, or vocabulary frequency. This is consistent with the view that higher visual cortex and natural language share partially overlapping representational constraints~\citep{doerig2025highlevel, schrimpf2021neural, wang2023better}. Our results suggest this is true of V4; V1, whose selectivity is dominated by sub-lexical features such as oriented edges and spatial frequency, shows weaker alignment with language embeddings, consistent with the expectation that the constraints shaping early visual cortex diverge from those shaping language.

However, the alignment we observe is partial, not complete. Neural activity correlates more strongly with visual embeddings ($r = 0.52$ in V4; $0.62$ in V1) than with language embeddings ($r = 0.36$ in V4; $0.51$ in V1), and hypothesis embeddings show the weakest alignment to neural activity ($r = 0.36$ in V4; $0.23$ in V1). This gradient likely reflects information loss across abstraction stages: captions omit fine-grained visual detail, while hypothesis generation further compresses within-class variability that may be neurally relevant. Crucially, text-to-image generation partially reverses this compression. Embeddings of hypothesis-generated images align with neural activity at $r = 0.43$ in V4 and $r = 0.34$ in V1, matching or exceeding the alignment achieved by hypothesis text alone; and neural responses to these generated images recover the full translate–hypothesize–generate loop at $r = 0.49$ in V4, though only $r = 0.25$ in V1, where low-level features such as orientation, spatial frequency, and contrast polarity are harder to capture through text-to-image generation. Whether this residual gap reflects limits of linguistic description, given its discrete vocabulary, or limitations of current vision-language models remains an open question.

The practical implication is that language embeddings can serve as a low-dimensional coordinate system for navigating neural selectivity space. Rather than characterizing each neuron in isolation, one can project neurons into a shared semantic space and identify functional clusters, interpolate between known tuning profiles, or predict which neurons should respond to novel feature combinations. This moves beyond one-neuron-at-a-time characterization toward a population-level cartography of selectivity, expressed in terms that are both human-interpretable and computationally tractable.

Finally, there is an evocative parallel with sparse coding~\citep{olshausen1996emergence}: the hypothesis-generation step uses reasoning to compress dense, redundant captions into short descriptions that separate activation from suppression. We present this as a conceptual analogy rather than a demonstrated result; the present work does not recover a population-level dictionary or characterize sparsity formally, and we do not claim that hypothesis paragraphs constitute a sparse semantic code in the technical sense.

\subsection{Challenges and limitations}
\label{sec:limitations}

\paragraph{Shared inductive biases across foundation models.}
Our pipeline depends on three large pretrained models whose training distributions overlap heavily with ImageNet: Gemini 3.0 Pro for image captioning and hypothesis synthesis, Imagen 4.0 for text-to-image generation, and a pretrained adversarially-trained ResNet50 backbone in the V4 digital twin. The screening corpus used to identify each neuron's most- and least-activating images is itself ImageNet. Strong verification performance on ImageNet-like generated stimuli may therefore partly reflect shared inductive biases across these models rather than genuine biological selectivity: features that are well-represented in the ImageNet distribution are simultaneously well-captioned by Gemini, well-rendered by Imagen, and well-modeled by the ResNet-based digital twin, and this mutual agreement inflates apparent closed-loop success without necessarily reflecting the neuron's tuning. The random-image baseline partially addresses this concern: random ImageNet samples pass through the same digital twin and inherit the same shared biases, yet drive far fewer neurons above the 95th percentile than hypothesis-generated images. Shared inductive biases alone therefore cannot account for the observed verification rates — the gap over the random baseline reflects selectivity-specific information carried by the hypothesis. What the baseline cannot rule out is whether that information generalizes beyond the ImageNet regime. Disentangling that would require replication on stimulus distributions outside the ImageNet training regime, such as scientific imagery, artistic renderings, or synthetic stimuli drawn from outside photographic conventions, and ideally with digital twins built on backbones pretrained on disjoint corpora. Until such replication is performed, the reported verification rates should be read as conditional on the ImageNet-centric ecosystem in which every stage of the pipeline was developed.

\paragraph{Prompt specificity and model limitations.}
Effective hypothesis generation required iterative refinement of the system prompt during pipeline development; per-neuron hypothesis generation itself is single-pass. This is a process that could be automated to an agentic workflow in the future. Vision-language models are known to anchor on semantic features, such as object identity and scene category, rather than the precise geometric or structural properties that often drive selectivity in early and mid-level visual cortex~\citep{tong2024eyeswideshut, yuksekgonul2023bagsofwords,kamath2023whatsup}.
This tendency is compounded by a well-documented limitation of multimodal
models: they struggle to perform comparative reasoning across many images
simultaneously, particularly when the shared feature is a low-level property
such as orientation or contour shape that does not map cleanly onto a lexical
category \citep{fu2024blink, kil2024compbench, meng2025mmiu}.
This limitation directly motivated our image-to-text translation stage: by
converting images into dense captions before synthesis, we shift the
cross-image comparison problem from vision to language, a regime in which
large language models are substantially more reliable. An additional benefit of operating in text is that hallucinations are suppressed: when the model reasons over captions rather than raw images, it finds similarities grounded in the actual descriptive content of the stimuli~\citep{zheng2023ddcot, leng2024vcd, zhou2024lure}.

\paragraph{In-silico verification.}
The verification loop currently closes against the digital twin rather than against biological recordings: hypothesis-generated images are tested by predicting responses with the encoding model, not by presenting stimuli to the animal and measuring spiking activity directly. While this is a limitation of our work, we have confidence in our results because we restrict analysis to well-predicted neurons (correlation-to-average $>0.4$) and because digital twins of comparable architecture have been validated against in-vivo responses in closed-loop stimulus design~\citep{Willeke2026-dq}. Nonetheless, direct experimental validation is an essential next step. Such experiments would also reveal whether our confidence thresholds are sufficient or whether additional neuron-selection criteria are needed.

\paragraph{Percentile-based verification vs.\ equivalence to original extremes.}
Our verification metric asks whether a hypothesis-generated image lands in an extreme percentile of the natural-image response distribution (e.g., above the 95th or below the 5th). This is a strong test against a random baseline, but it is a weaker claim than equivalence to the specific top-15 MAIs or bottom-15 LAIs used to derive the hypothesis: the latter occupy a small subregion within the extreme tails, and generated images that land in the same tail may still fall short of the activation levels reached by the defining extremes. Directly comparing generated-image responses to the actual MAI and LAI responses per neuron would provide a stronger test and is a natural next step. The present results should therefore be read as establishing that hypotheses drive neurons into the extreme response regime, rather than that they recreate the original extremes exactly.

\paragraph{Spatial reasoning in language models.}
Semantic hypotheses reliably captured \textit{what} features drive a neuron but often failed to specify \textit{where} those features should appear, at what scale, or at what orientation. This reflects a well-documented weakness of vision-language models in precise spatial reasoning \citep{Kamath2023-qr, Chen2024-sz, Stogiannidis2025-lq}, and it motivated the affine augmentation stage. To verify that this spatial search did not artificially inflate verification scores, we applied identical augmentations to random images and confirmed that response gains are substantially smaller. Notably, we observed qualitative improvements in spatial descriptions between model generations, suggesting that continued progress in foundation models will reduce reliance on post-hoc spatial correction.

\paragraph{Asymmetry between excitatory and suppressive hypotheses in V1.} Semantic hypotheses verified less reliably at the suppressive pole in V1. We can only speculate about the origin of this asymmetry between low- and high-activation regimes, but several non-exclusive explanations seem plausible. One possibility is that a single semantic hypothesis collapses a potentially high-dimensional suppressive manifold, the full equivalence class of stimuli that fail to match preferred tuning along any of several axes, into one description. We did not systematically probe these suppressive invariances, and doing so may be necessary to recover the full structure of what does \emph{not} drive a neuron. A second possibility is a current limit on linguistically describing sub-lexical V1 features such as precise orientation and spatial frequency: text-to-image models may render such features only approximately, and imprecise rendering may still land within the broad suppressive manifold, whereas at the excitatory pole the same imprecision would miss the much narrower preferred region. We cannot localize this bottleneck from these experiments alone since it may reflect that natural language rarely specifies such features in training corpora, that current vision-language and text-to-image models do not reliably encode or render them even when prompted, that language as a system lacks the granularity these features require, or some combination of the three. Disentangling these possibilities will require targeted experiments and may narrow as foundation models improve on fine-grained spatial specification.

\subsection{Automated characterization as a stepping stone toward agentic neural discovery}

Our framework can be viewed as an initial instance of a broader paradigm for studying complex systems at scale: constructing high-fidelity ``digital twin'' models from large-scale data and automatically distilling their behavior into human-interpretable descriptions. In the present work, this pipeline operates in a fixed, open-loop manner: each stage runs once per neuron without feedback between
iterations. This is a deliberate trade-off, since deterministic execution scales to thousands of neurons and yields reproducible characterizations, but it stops short of the closed-loop paradigm that agentic systems afford, where experimental outcomes drive decisions about what to probe next \citep[e.g.][]{shaham2024multimodal}. Extending our approach toward true agentic operation is a natural and tractable next step.

Agentic exploration requires cheap inner-loop experiments; functional digital twins of the brain provide exactly that. A validated neural encoding model lets an agent query predicted responses to arbitrary stimuli at negligible cost, running hypothesis generation and revision entirely in-silico before any biological experiment is needed. Sufficiently confident predictions would then be handed to experimentalists for confirmation, creating a productive division of labor between scalable hypothesis generation and targeted empirical testing. More broadly, this suggests a general framework for scientific discovery in complex domains: large-scale datasets are used to construct expressive models of a system, these models are distilled into interpretable components, and agentic systems use these representations to guide adaptive experimentation. Rather than replacing human scientists, such systems may reallocate effort: automating the exhaustive exploration of high-dimensional hypothesis spaces while preserving human judgment in selecting, validating, and contextualizing discoveries.

Whether agentic systems can surface genuinely novel discoveries, rather than accelerating human-directed inquiry, remains an open and important question \citep{lu2024ai}. For systems neuroscience, the most defensible near-term vision is a collaborative one: automated and ultimately agentic tools that compress large-scale neural characterization and deliver prioritized, falsifiable hypotheses, while researchers retain judgment over what constitutes a discovery. The framework presented here is a concrete step toward that vision.

\section{Methods}
\label{sec:methods}

Our framework translates the selectivity of individual neurons into interpretable semantic hypotheses through a three-stage pipeline (Figure~\ref{fig:overview}). First, we convert a large corpus of natural images into dense textual descriptions using a vision-language model. Second, we leverage functional digital twin models of macaque visual cortex to identify images at both extremes of each neuron's activation spectrum, both those that maximally excite and those that maximally suppress, and synthesize their descriptions into semantic hypotheses characterizing the neuron's preferred and non-preferred features. Third, we validate these hypotheses by generating novel images from the semantic descriptions and testing whether they elicit the predicted neural responses in closed-loop experiments on the digital twin. This pipeline ensures that each semantic interpretation is not merely a post-hoc rationalization but a generative model that makes falsifiable predictions about neural activity.

\subsection{Image-to-Text Translation}

The foundation of our approach is a dense captioning procedure that converts visual content into detailed textual descriptions (Figure~\ref{fig:translate}). We use Gemini 3.0 Pro to generate exhaustive descriptions of images from ImageNet, prompting the model to describe visual content with sufficient precision that the image could, in principle, be reconstructed from the text alone.

\begin{tcolorbox}[
    enhanced,
    colback=magenta!3!white,
    colframe=magenta!70!black,
    boxrule=1pt,
    arc=3mm,
    left=10pt,
    right=10pt,
    top=8pt,
    bottom=8pt,
    fontupper=\small,
]
\textbf{\textcolor{magenta!70!black}{Captioning Prompt:}}
\vspace{3pt}

\textit{You are given a single image that is partially masked by a black circular border. Completely ignore the mask and only describe what is visible inside the circle. Write exactly one paragraph that describes the contents with exhaustive visual detail, as if someone must recreate the image from your words alone. Be extremely concrete: describe colors with precise shades, materials and textures, the exact spatial relationships between objects (including their relative positions, distances, and overlaps), their precise orientation, lighting conditions, shadows, and with exact proportions between objects. Scale the proportions of object in reference to the overall image size, including the border, however do not mention the circular boundary in the description. Start with dominant elements and progress to finer details. If elements are partially cut off by the boundary, describe only what is visible. Avoid interpretation, speculation, or mentioning the mask. Ensure correctness of spatial location and orientation. Do not use markdown or latex. Conclude writing only when your paragraph is sufficient to describe the entire image. Respond right away with a single paragraph - no introductory text or explanations.}
\end{tcolorbox}

This captioning procedure departs from standard image captioning in its emphasis on visual fidelity over semantic summarization. Rather than producing terse labels (e.g., ``a dog in a park''), the model generates multi-sentence descriptions specifying precise color shades, surface textures, spatial arrangements, lighting conditions, and the proportional relationships between objects. These dense captions preserve the fine-grained visual information necessary for subsequent hypothesis generation, while recasting the problem from one modality to another: from vision to language.

To validate the faithfulness of this translation, we implement a round-trip reconstruction test. For each original image, we use its generated caption to synthesize a new image via a text-to-image model (Imagen 4.0), then measure the correspondence between original and reconstructed images in DINOv3 embedding space. We compute cosine similarity between DINOv3 embeddings, defined by the CLS token of the image, of paired images and compare against a baseline of similarities to randomly sampled ImageNet images. Reconstructed images consistently exhibit higher similarity to their source images than to unrelated images (Figure~\ref{fig:translate}), confirming that captions preserve visually relevant information through the translation process.

\subsection{Functional Digital Twin Models}

To characterize neural selectivity across a stimulus space far larger than any single recording session can survey, we employ functional digital twin models, deep learning models trained to predict the responses of individual biological neurons to arbitrary images. These models serve as differentiable surrogates for the recorded neural populations, enabling large-scale screening and closed-loop experimentation in-silico.

Each digital twin combines a shared convolutional neural network (CNN) core with neuron-specific readout layers. Following prior work demonstrating that successive stages of the visual hierarchy align with progressively deeper network layers~\citep{Cadena2023-qb, Fu2024-og}, we use distinct architectures for different cortical areas. For V1 neurons, we employ the first layer of a ConvNeXt architecture~\citep{Woo2023-wb}, fine-tuning the convolutional core on recorded responses while learning neuron-specific Gaussian readouts~\citep{Lurz2022-gu}. For V4 neurons, we use a pretrained adversarially-trained ResNet50~\citep{Salman2019-aa} as a fixed feature extractor, training linear readouts on top of layer~3 representations~\citep{Willeke2023-ax}.

Model performance is evaluated on held-out test images using the Pearson correlation between predicted and observed responses averaged across stimulus repetitions. We retain only neurons achieving a correlation-to-average above 0.4, yielding high-confidence digital twins for 438 V1 neurons (84\% of the recorded population) and 205 V4 neurons (52\% of the recorded population). These thresholds increase our confidence that subsequent analyses reflect genuine neural selectivity rather than model artifacts.

\subsection{Identification of Most and Least Activating Images}

Recent work has demonstrated that many neurons in visual cortex maintain elevated baseline firing rates and exhibit \emph{dual-feature selectivity}: they are not only driven by preferred stimuli but are also systematically suppressed by distinct non-preferred features~\citep{Franke2026-qb}. To capture the full dynamic range of this bidirectional modulation, we characterize both ends of each neuron's activation spectrum.

For each neuron, we screen 1.2 million ImageNet images through the digital twin to identify the \emph{most activating images} (MAIs), natural stimuli predicted to elicit the strongest responses, and the \emph{least activating images} (LAIs), those predicted to suppress activity most strongly below baseline. All images are normalized to identical $\ell_2$ norms within each neuron's receptive field prior to screening, ensuring that response differences reflect feature selectivity rather than contrast variation. The resulting distributions of predicted responses across all images characterize each neuron's selectivity profile, with MAIs and LAIs occupying opposite tails (Figure~\ref{fig:hypothesis}).

This dual characterization is essential for non-sparse neurons, whose graded, approximately linear responses to natural images are jointly shaped by similarity to both their preferred and non-preferred features~\citep{Franke2026-qb}. Restricting the analysis to MAIs alone would capture only half of the encoding axis along which these neurons operate.

\subsection{Semantic Hypothesis Generation}

The core of the interpretability pipeline is the synthesis of semantic hypotheses from the images occupying both extremes of the activation spectrum. For each neuron, we retrieve the dense captions corresponding to its top 15 most activating images and its bottom 15 least activating images, then prompt Gemini 3.0 Pro to identify the visual features that distinguish these two sets and could therefore explain the neuron's selectivity.

The hypothesis generation prompt instructs the model to characterize: (1)~\emph{Excitatory selectivity}, the core visual features, shapes, textures, or semantic motifs that drive strong activation, described in concrete visual terms rather than abstract categories (e.g., ``a glossy, dark spherical form with a bright specular highlight'' rather than ``an eye''); (2)~\emph{Suppressive selectivity}, the distinct visual features associated with minimal activation, (3)~\emph{Invariances}, visual properties that can vary without substantially altering the response, such as object identity, material, lighting, or viewpoint; (4)~\emph{Spatial requirements}, how activating and suppressive features are arranged within the image, including their relative size, placement, and orientation; (5)~\emph{Color and contrast}, dominant hues of activating features, typical background tones, and consistent polarity relationships; and (6)~\emph{Quantitative spatial detail}, approximate size and position of the critical features within the frame. The full prompt is provided in Appendix~\ref{prompt:hypothesis}.

This approach exploits a key asymmetry in current foundation models: while vision-language models struggle to reason jointly across many images presented simultaneously, large language models excel at multi-document synthesis from text. By first translating images into language, we convert cross-image visual comparison, a task that strains current multimodal architectures, into textual pattern extraction, a regime where language models are powerful and reliable.

\subsection{Hypothesis Verification via Generative Synthesis}

A semantic hypothesis is only as credible as the predictions it licenses. We validate each hypothesis through a closed-loop procedure: converting the semantic description into prompts for a text-to-image model, generating novel images, and testing whether these images modulate the neuron's activity in the direction predicted by the hypothesis.

\paragraph{Image generation from hypotheses.}
From each semantic hypothesis, we use Gemini 3.0 Pro to generate 15 diverse image prompts that embody the hypothesized excitatory selectivity while varying along the identified invariance dimensions. For instance, if a hypothesis specifies selectivity for ``dark rounded objects with specular highlights against light, desaturated backgrounds,'' prompts describe this motif instantiated across different object categories, materials, and scenes, always preserving the core selectivity features while varying irrelevant properties. We similarly generate 15 prompts targeting the hypothesized suppressive features. Each prompt is passed to a text-to-image model (Imagen 4.0) to produce candidate stimuli. The full prompt template is provided in Appendix~\ref{prompt:diffusion}.

\paragraph{Affine augmentation of generated images.}

Natural images that strongly activate a neuron do so across a range of minor spatial configurations, such as small shifts in position, rotations, and modest changes in scale, that fall within the tolerance of the neuron's receptive field. To ensure that our verification is robust to the particular spatial layout chosen by the text-to-image model, and to search for configurations that may better align with each neuron's spatial preferences, we apply a battery of affine transformations to every generated image. Specifically, each image is augmented with compositions of (i)~full rotations sampled uniformly from $[0^{\circ}, 360^{\circ})$, (ii)~translations of up to 1.5\% of the image extent along each axis, and (iii)~isotropic scalings between $0.9\times$ and $1.1\times$, with bilinear interpolation and reflection padding at the boundaries. We use full rotations because V1 neurons exhibit sharp orientation tuning, and the text-to-image model has no mechanism to generate stimuli at a neuron's preferred orientation; a complete rotational search is therefore necessary to align the generated edge or grating with the neuron's tuning. We apply the same full-rotation augmentation to V4 neurons to maintain a fair and uniform procedure across areas,  though the results suggest V4 selectivity is generally less orientation-dependent. For each generated image, we produce a set of augmented variants, pass all variants through the digital twin, and retain the transformation yielding the highest (or, for suppressive hypotheses, lowest) predicted response. This augmentation strategy serves as a controlled spatial search that respects the neuron's receptive field geometry without altering the semantic content of the image.

\paragraph{Quantifying verification success.}
We quantify the effectiveness of each hypothesis by computing the response percentile of the best-augmented generated images relative to the full distribution of predicted responses across the 1.2 million ImageNet screening images. If the semantic hypothesis accurately captures the neuron's excitatory selectivity, generated images should rank in the upper tail of this distribution; if it captures suppressive selectivity, the corresponding images should fall in the lower tail.

To establish that any observed activation is attributable to the semantic content of the hypothesis rather than to incidental properties of generated images or the augmentation procedure itself, we apply the same affine augmentation pipeline to a set of randomly selected natural images that bear no specific relationship to the neuron's selectivity (Figure~\ref{fig:verification}). We search over the same augmentation space, retaining the transformation that maximizes (or minimizes) the predicted response for each random image. This control isolates the contribution of spatial search alone, stripped of any semantic targeting, and yields response percentiles that cluster near the median of the screening distribution, demonstrating that the high (or low) predicted responses to hypothesis-driven images arise from the match between their semantic content and the neuron's feature preferences, rather than from artifacts of the generation or augmentation process.

\subsection{Representational Similarity Analysis}
\label{sec:rsa_methods}

For each cortical area, we constructed neuron-by-neuron representational similarity matrices (RSMs) in six spaces: neural population activity, DINOv3 image embeddings, Qwen3 0.6B caption embeddings, Qwen3 0.6B hypothesis embeddings, DINOv3 embeddings of hypothesis-generated images, and neural responses to those generated images. In each space, a neuron is represented by the mean embedding (or activity vector) over a set of its most-activating images, and pairwise neuron similarity is the Pearson correlation between the resulting per-neuron vectors. Cross-space alignment was measured by the Pearson correlation between the upper triangles of two RSMs, following standard RSA \citep{kriegeskorte2008representational}.

\paragraph{Held-out construction.} Ordering an RSM by neural similarity and then reading alignment off another RSM in the same ordering risks inheriting structure from the ordering itself rather than from genuine cross-space correspondence. To avoid this, we split each neuron's 15 most-activating images into a 7-image held-out set and an 8-image evaluation set. The held-out set was used \emph{only} to compute neural similarities between neurons, which defined the row/column ordering of all subsequently displayed RSMs. All reported cross-space correlations and the matrices in Figure~\ref{fig:alignment} were then computed on the disjoint 8-image evaluation set, with each neuron represented by the mean embedding (or activity vector) of those 8 images in each of the six spaces. Any structure preserved across spaces under this construction must reflect genuine cross-modal correspondence rather than circular reuse of the clustering signal.

\paragraph{Permutation test.} For each pair of spaces, we tested whether the observed RSA correlation exceeded what would be expected under a permuted-label null. We jointly permuted the row and column labels of one of the two RSMs (preserving its internal geometric structure, as recommended by \citealp{kriegeskorte2013representational}), recomputed the upper-triangle correlation, and repeated this $10{,}000$ times to build an empirical null distribution. We report the observed correlation, the one-sided $p$-value (fraction of null correlations $\geq$ observed), and the 95th percentile of the null. Permuting individual upper-triangle entries was avoided, as it destroys the matrix structure and produces an implausibly easy null. Full per-pair statistics for both V1 and V4 are given in Appendix Table~\ref{tab:rsa_perm}.

\subsection*{Author Contributions}
\textbf{VL}: Conceptualization, Methodology, Software, Validation, Formal analysis, Investigation, Data Curation, Writing - Original Draft, Visualization, Project Administration
\textbf{KF}: Conceptualization, Methodology, Investigation, Data Curation, Writing - Original Draft, Supervision, Project Administration
\textbf{TRS}: Conceptualization, Methodology, Writing - Review \& Editing
\textbf{SG}: Conceptualization, Methodology, Writing - Review \& Editing
\textbf{AT}: Conceptualization, Supervision, Funding Acquisition, Writing - Review \& Editing
\textbf{SS}: Conceptualization, Supervision, Funding Acquisition, Writing - Review \& Editing
\textbf{NK}: Conceptualization, Methodology, Software, Investigation, Data Curation, Writing - Original Draft, Supervision, Project Administration

\subsection*{Acknowledgment}

This work was supported by the James Fickel Enigma Project Fund awarded to AST and SS. We thank Paul Steffan and Edgar Walker for preliminary exploration of related ideas, as well as useful discussions with Dan O'Shea, Mahdi Ramadan, Alex Gilbert, Liam Storan, Hasan Atakan Bedel.

\bibliographystyle{unsrtnat}
\bibliography{references}
\appendix
\clearpage
\section{Appendix}

\subsection{RSA permutation-test statistics}
\label{app:rsa_perm}

Table~\ref{tab:rsa_perm} reports the full set of cross-space RSA correlations for V1 and V4, together with one-sided permutation $p$-values and the 95th percentile of the corresponding label-permutation null (10{,}000 permutations; see Methods~\ref{sec:rsa_methods}). All correlations are computed on the held-out 8-image evaluation set, with the row/column ordering defined by neural clustering on the disjoint 7-image set.

\begin{table}[h]
\centering
\small
\setlength{\tabcolsep}{6pt}
\renewcommand{\arraystretch}{1.05}
\begin{tabular}{lrrrrrr}
\toprule
& \multicolumn{3}{c}{V1} & \multicolumn{3}{c}{V4} \\
\cmidrule(lr){2-4} \cmidrule(lr){5-7}
Pair & $r$ & $p$ & null 95\% & $r$ & $p$ & null 95\% \\
\midrule
Caption vs Hypothesis                & $+0.377$ & $<\!10^{-4}$ & $+0.006$ & $+0.357$ & $<\!10^{-4}$ & $+0.012$ \\
Caption vs Neural                    & $+0.508$ & $<\!10^{-4}$ & $+0.020$ & $+0.356$ & $<\!10^{-4}$ & $+0.012$ \\
Caption vs Image                     & $+0.674$ & $<\!10^{-4}$ & $+0.023$ & $+0.670$ & $<\!10^{-4}$ & $+0.012$ \\
Caption vs Image (hyp-gen)           & $+0.372$ & $<\!10^{-4}$ & $+0.005$ & $+0.361$ & $<\!10^{-4}$ & $+0.012$ \\
Caption vs Neural (hyp-gen)          & $+0.146$ & $<\!10^{-4}$ & $+0.020$ & $+0.192$ & $<\!10^{-4}$ & $+0.011$ \\
Hypothesis vs Neural                 & $+0.232$ & $<\!10^{-4}$ & $+0.006$ & $+0.358$ & $<\!10^{-4}$ & $+0.012$ \\
Hypothesis vs Image                  & $+0.324$ & $<\!10^{-4}$ & $+0.006$ & $+0.369$ & $<\!10^{-4}$ & $+0.012$ \\
Hypothesis vs Image (hyp-gen)        & $+0.600$ & $<\!10^{-4}$ & $+0.006$ & $+0.600$ & $<\!10^{-4}$ & $+0.012$ \\
Hypothesis vs Neural (hyp-gen)       & $+0.259$ & $<\!10^{-4}$ & $+0.006$ & $+0.365$ & $<\!10^{-4}$ & $+0.010$ \\
Neural vs Image                      & $+0.615$ & $<\!10^{-4}$ & $+0.032$ & $+0.522$ & $<\!10^{-4}$ & $+0.014$ \\
Neural vs Image (hyp-gen)            & $+0.338$ & $<\!10^{-4}$ & $+0.005$ & $+0.429$ & $<\!10^{-4}$ & $+0.012$ \\
Neural vs Neural (hyp-gen)           & $+0.249$ & $<\!10^{-4}$ & $+0.028$ & $+0.491$ & $<\!10^{-4}$ & $+0.020$ \\
Image vs Image (hyp-gen)             & $+0.346$ & $<\!10^{-4}$ & $+0.004$ & $+0.454$ & $<\!10^{-4}$ & $+0.012$ \\
Image vs Neural (hyp-gen)            & $+0.102$ & $<\!10^{-4}$ & $+0.031$ & $+0.267$ & $<\!10^{-4}$ & $+0.017$ \\
Image (hyp-gen) vs Neural (hyp-gen)  & $+0.452$ & $<\!10^{-4}$ & $+0.005$ & $+0.447$ & $<\!10^{-4}$ & $+0.011$ \\
\bottomrule
\end{tabular}
\caption{Per-pair RSA correlations, permutation $p$-values, and 95th-percentile null thresholds for V1 and V4. Hypothesis generation (\textit{hyp-gen}) refers to images synthesized from semantic hypotheses via Imagen 4.0 and the corresponding neural responses to those images. Every observed correlation exceeds its label-permutation null by a wide margin.}
\label{tab:rsa_perm}
\end{table}
\subsection{Generating Image Captions}
\label{prompt:captioning}
We prompt Gemini 3.0 Pro with each image to produce a single dense paragraph describing its visual content with sufficient precision that the image could, in principle, be reconstructed from the text alone.
\begin{tcolorbox}[
    enhanced,
    breakable,
    colback=magenta!3!white,
    colframe=magenta!70!black,
    boxrule=1pt,
    arc=3mm,
    left=10pt,
    right=10pt,
    top=8pt,
    bottom=8pt,
    fontupper=\small,
]
\textbf{\textcolor{magenta!70!black}{Captioning Prompt:}}
\vspace{3pt}

\textit{You are given a single image that is partially masked by a black circular border. Completely ignore the mask and only describe what is visible inside the circle. Write exactly one paragraph that describes the contents with exhaustive visual detail, as if someone must recreate the image from your words alone. Be extremely concrete: describe colors with precise shades, materials and textures, the exact spatial relationships between objects (including their relative positions, distances, and overlaps), their precise orientation, lighting conditions, shadows, and with exact proportions between objects. Scale the proportions of object in reference to the overall image size, including the border, however do not mention the circular boundary in the description. Start with dominant elements and progress to finer details. If elements are partially cut off by the boundary, describe only what is visible. Avoid interpretation, speculation, or mentioning the mask. Ensure correctness of spatial location and orientation. Do not use markdown or latex. Conclude writing only when your paragraph is sufficient to describe the entire image. Respond right away with a single paragraph - no introductory text or explanations.}
\end{tcolorbox}

\subsection{Generating Neuron Selectivity Hypotheses}
To convert neural selectivity patterns into interpretable semantic descriptions, we prompt a large language model to synthesize detailed captions of highly activating images into a concise neuron selectivity hypothesis (Figure~\ref{fig:hypothesis}).
\begin{tcolorbox}[
    enhanced,
    colback=violet!3!white,
    colframe=violet!70!black,
    boxrule=1pt,
    arc=3mm,
    left=10pt,
    right=10pt,
    top=8pt,
    bottom=8pt,
    fontupper=\small
]
\textbf{\textcolor{violet!70!black}{Hypothesis Generation Prompt:}}
\vspace{3pt}

\texttt{\{\{CONTEXT\_BLOCK\}\}} \\
\textit{You are given a set of detailed captions describing images that strongly activate a single neuron in a vision model. Your task is to synthesize these captions into a single, coherent neuron selectivity hypothesis that captures the following aspects:}

\vspace{3pt}
\begin{enumerate}[leftmargin=*, itemsep=2pt, topsep=0pt]
    \item \textbf{Selectivity:} Identify the core, consistent visual features, shapes, textures, or semantic concepts that drive the neuron's activation. Describe these features concretely in visual terms rather than abstract categories (e.g., ``a thin curved metallic edge'' rather than ``a tool'').

    \item \textbf{Invariances:} Explain which visual properties can vary (e.g., object identity, material, color tone, lighting, background, scale, or perspective) without substantially reducing activation.

    \item \textbf{Spatial and dimensional requirements:} Describe how the activating features are spatially arranged or proportioned within the image—include their relative size, placement, and orientation.

    \item \textbf{Color and contrast:} Explicitly describe: the dominant colors or hues of the activating features; the typical colors or tones of the background; whether there is a consistent contrast or polarity between the feature and its background.

    \item \textbf{Quantitative spatial detail:} Include one explicit sentence specifying the approximate relative size and position of the activating feature(s) within the frame (e.g., ``The feature typically occupies 25–50\% of the image height and is centered or slightly to the lower left.'').

    \item \textbf{Synthesis and conciseness:} Summarize your findings into one concise, cohesive paragraph that captures the neuron's essential selectivity, invariances, spatial configuration, and color relationships. Avoid irrelevant scene details or exhaustive lists; emphasize the key visual pattern that consistently drives activation.
\end{enumerate}

\vspace{3pt}
\textit{Style guideline: Write in a precise, neutral, and observational tone similar to the following example:}

\vspace{2pt}
\begin{quote}
\small
The neuron responds strongly to a dark colored object—typically black or dark brown—set against a light, desaturated background of pale beige, off-white, or light pink. The activating feature has an elongated or rounded shape and is centered or slightly off-center, occupying about 20–60\% of the image. The neuron remains invariant to the object's identity, texture, precise hue, lighting, and background details, as long as the strong dark–light contrast and overall form are preserved.
\end{quote}
\label{prompt:hypothesis}
\end{tcolorbox}

\subsection{Generating Images from Neuron Hypotheses}

To verify neuron selectivity hypotheses, we generate synthetic images that should activate the neuron according to its predicted preferences. We prompt a large language model to create detailed image descriptions that embody the hypothesis while introducing controlled variations (Figure~\ref{fig:verification}).

\begin{tcolorbox}[
    enhanced,
    colback=blue!3!white,
    colframe=blue!70!black,
    boxrule=1pt,
    arc=3mm,
    left=10pt,
    right=10pt,
    top=8pt,
    bottom=8pt,
    fontupper=\small
]
\textbf{\textcolor{blue!70!black}{Image Generation Prompt:}}
\vspace{3pt}

\textit{Neuron Hypothesis:}
\texttt{\{\{CONTEXT\_BLOCK\}\}}

\vspace{3pt}
\textit{You are given a neuron hypothesis above that strongly activates a neuron in a vision model. Your task is to produce 15 different vividly detailed descriptions of new images that would be described by the hypothesis.}

\vspace{3pt}
\textit{Each generated description should:}

\vspace{2pt}
• Explicitly describe a new image that embodies the neuron hypothesis using natural language (not numbers or coordinates) to specify colors, spatial arrangements, orientations, and scales.

\vspace{2pt}
• Introduce variations following the invariances but not deviate from the size and color of the feature as described by the hypothesis. If a size range is given in the hypothesis (x\%--y\%), ensure the description is only of the lowest end of the range (x\%).

\vspace{2pt}
• Along with spatial constraints, each image must strictly follow the color of the feature as well as the color of the background as described in the hypothesis.

\vspace{2pt}
• Be richly descriptive.

\vspace{3pt}
\textit{Here is an example of a detailed description:}

\vspace{2pt}
\begin{quote}
\footnotesize
A vibrant Rainbow Lorikeet, oriented facing slightly to the right with its head tilted slightly downwards, occupies the central-left portion of the visible frame. Its crown, nape, and chin are a deep, rich indigo blue, almost violet, which transitions to a vibrant emerald green on the cheeks, extending from just below the eye to the back of the head. The eye is a small, dark, round orb with a reddish-orange iris and a black pupil, positioned slightly above the horizontal midline of the head, encircled by a thin, dark ring. The beak is a striking, bright vermilion red, conical in shape with a slight downward curve at the tip of the upper mandible, which slightly overlaps the lower; a small, dark nostril is visible at the base of the upper mandible. Numerous tiny, clear water droplets cling to the feathers on the bird's head and back, reflecting light. Below the blue chin, a narrow band of bright yellow-green forms a collar, separating the head from the chest. The chest is a brilliant, fiery orange, transitioning to a deep scarlet red in its central portion, with individual feathers clearly delineated, giving a slightly ruffled texture. This vibrant chest plumage covers the upper torso. Below the orange chest, the feathers on the abdomen and sides are a rich, dark indigo blue, similar to the head, with some emerald green feathers visible on the lower sides. The visible portion of the bird's back and folded wing is a lush emerald green, also adorned with small water droplets. The bird is well-lit from the front-left, creating subtle highlights on the feathers and a soft shadow beneath its beak. A rough, natural wooden branch, approximately one-fifth the height of the bird and one-tenth its body width in diameter, extends diagonally from the lower-right corner of the visible area upwards and slightly left, passing behind the bird's lower body. Its surface is textured with visible bark patterns, grooves, and irregularities, in muted shades of brown and grey. The background is a heavily blurred, soft expanse of muted olive green and darker forest green, with some indistinct brown-grey patches, suggesting out-of-focus foliage or distant natural elements, providing a shallow depth of field that places the bird sharply in focus against the soft backdrop.
\end{quote}
\label{prompt:diffusion}
\end{tcolorbox}

\noindent These richly detailed descriptions are then passed to a text-to-image model (Imagen 4.0) to synthesize images for testing against the digital twin model, completing the closed-loop verification of neuron selectivity.

\subsection{Generative Model Parameters}
\begin{table}[h]
\centering
\small
\label{tab:genhparams}
\begin{tabular}{@{}lll@{}}
\toprule
\textbf{Parameter} & \textbf{Value} & \textbf{Notes} \\
\midrule
\multicolumn{3}{l}{\textit{Gemini 3 Pro (text/multimodal generation)}} \\
\midrule
Model ID            & \texttt{gemini-3-pro-preview} & via Google GenAI SDK / Vertex AI \\
Temperature (online)& $1.0$ &  captioning, hypothesis generation, activation-prompt generation (Default) \\
Top-$p$             & API default (0.95) & not overridden \\
Top-$k$             & API default (64) & not overridden \\
Max output tokens   & API default & not overridden \\
Thought summaries   & enabled (batch) & \texttt{include\_thoughts=true} \\
Response format     & \texttt{application/json} & Pydantic-typed structured output \\
Safety filters      & \texttt{BLOCK\_NONE} (all 4 categories) & dangerous, harassment, hate, sexual \\
\midrule
\multicolumn{3}{l}{\textit{Imagen 4 (text-to-image generation)}} \\
\midrule
Model ID             & \texttt{imagen-4.0-generate-001} & via Vertex AI (\texttt{api\_version=v1}) \\
Images per prompt    & $1$ & one sample per activation prompt \\
Aspect ratio         & $1{:}1$ & square output \\
Negative prompt      & \multicolumn{2}{l}{``text, numbers, annotations, labels, percentages, measurements, overlays''} \\
Person generation    & \texttt{ALLOW\_ALL} & \\
Safety filter level  & \texttt{block\_only\_high} & \\
Watermark            & disabled & \texttt{add\_watermark=false} \\
\bottomrule
\end{tabular}
\caption{Generative model hyperparameters used in our pipeline. Parameters
not listed in the configuration are inherited from the provider's API
defaults; we report this explicitly for transparency.}
\end{table}
Except for experiments invoking the Gemini API, all models (including the digital twins) fit on a single NVIDIA GeForce RTX 3090 GPU. Analyses were performed on a workstation with 100 GB of RAM, although substantially smaller configurations would suffice. On this hardware, all GPU-bound computations completed within less than 1 hour of wall-clock time, whereas the most demanding CPU-bound analyses required up to 10 hours of wall-clock time.

\end{document}